\documentclass[10pt,english]{article}
\usepackage[T1]{fontenc}
\usepackage[latin1]{inputenc}
\usepackage{array}
\usepackage{subfigure}
\usepackage{graphicx}
\usepackage{amsmath}
\usepackage{float}
\usepackage{color}
\usepackage{epsfig}

\usepackage{algorithm}
\usepackage[noend]{algorithmic}

\makeatletter

\providecommand{\tabularnewline}{\\}

\usepackage{subfigure}

\usepackage{babel}
\makeatother

\title{Discovering Patterns in Multi-neuronal Spike Trains using the Frequent Episode Method}

\author{
K.P.Unnikrishnan\footnote{General Motors R\&D Center, Warren, MI} and 
Debprakash Patnaik\footnote{Dept. Elecetrical Engineering, Indian Institute of Science, 
Bangalore} and 
P.S.Sastry\footnote{Dept. Elecetrical Engineering, Indian Institute of Science, 
Bangalore}
}
\date{}

\begin{document}

\baselineskip 16pt

\maketitle

\thispagestyle{empty}

\section*{ABSTRACT}
Discovering the 'Neural Code' from multi-neuronal spike trains is an important task in 
neuroscience. For such an analysis, it is important to unearth interesting regularities 
in the spiking patterns. In this report, we present an efficient method for automatically 
discovering synchrony, synfire chains, and more general sequences of neuronal firings. 
We use the Frequent Episode Discovery framework of Laxman, Sastry, and Unnikrishnan (2005), 
in which the episodes are represented and recognized using finite-state automata. Many 
aspects of functional connectivity between neuronal populations can be inferred from the 
episodes. We demonstrate these using simulated multi-neuronal data from a Poisson model. 
We also present a method to assess the statistical significance of the discovered episodes. 
Since the Temporal Data Mining (TDM) methods used in this report can analyze data from 
hundreds and potentially thousands of neurons, we argue that this framework is appropriate 
for discovering the `Neural Code'.

\section{INTRODUCTION}
\label{sec:intro}

Analyzing spike trains from hundreds of neurons is an important and exciting problem. 
By using experimental techniques such as Micro Electrode Arrays or imaging of 
neural currents through voltage-sensitive dyes etc., spike data can be recorded 
simultaneously from many neurons \cite{Ikegaya2004,Potter2006}.  
Automatically discovering patterns (regularities) in these spike trains can lead to better 
understanding of the functional relationships within the system that produced the spikes. 
Such understanding of functional relations embedded in spike trains lead to many 
applications, e.g., better brain-machine interfaces. Such an analysis 
can also ultimately allow us to systematically answer the question, "is there a neural code?".


In this paper, we present some novel methods to analyze spike train data,  
 based on the method of frequent episode discovery in time-ordered event 
sequences \cite{Mannila1997,Srivats2005,Deb2006}, which is from the field of temporal data 
mining. 
Temporal data mining is concerned with analysis of large sequential data
sets \cite{Srivats-survey2005}. Such data sets with temporal dependencies
frequently occur in
many business, engineering and scientific scenarios.
Frequent episode discovery, originally proposed in \cite{Mannila1997},
is one of the popular frameworks in temporal data mining.
Here, the data is viewed as a time-ordered sequence of events 
where each event is characterized by an event type and a time of occurrance. 
A few examples of such data are alarms in a telecommunication network, 
fault logs of a manufacturing plant etc. 
The goal of the analysis is to
unearth temporal patterns (called episodes) that occur sufficiently often along
that sequence. These discovered patterns are called frequent episodes. 
The multi-neuronal spike train data is also 
a sequential or time-ordered data stream of events where each event is a spike at 
a particular time and  
 the event type would be the neuron (or the electrode in the 
micro electrode array) that generated the spike. 
 Since functionally
interconnected neurons tend to fire in certain precise patterns, discovering
frequent patterns in such temporal data can help understand the underlying
neural circuitry. In this paper, we argue that the frequent episodes framework is ideally
suited for such analysis.
There are efficient algorithms for 
automatically detecting many types of frequent episodes \cite{Mannila1997,Srivats2005}. 
However, as we shall see, in analyzing neural spiking data, one needs methods that can 
discover frequent episodes under different kinds of temporal constraints. We explain some 
datamining algorithms for frequent episode discovery under such temporal 
constraints \cite{Deb2006}. Through extensive simulation studies using both synthetic 
and real neural data, we argue that the frequent episodes framework is ideally suited 
for this application. We show that these datamining techniques provide 
 a very efficient and general purpose methodology for detecting many types 
of interesting patterns in spike data. 

Most of the currently available methods for analyzing spike train data 
 rely on quantities that can be
computed through cross correlations among spike trains (time shifted with respect to
one another) to identify interesting patterns in spiking activity. There are
methods to look for specific patterns and assess their statistical significance
under a null hypothesis that different spike trains are \textit{iid} Bernoulli
processes \cite{Abeles1988,Lee2004,Tetko2001}. Most such methods can not look for
patterns that involve more than 3 or 4 neurons due to the ubiquitous curse
of dimensionality. Looking for repeated 
occurrences of patterns of firing involving many neurons becomes infeasible due to the 
combinatorial explosion of candidate patterns that one should look for. The data mining 
approach  tackles this by adopting the same basic idea as in the Apriori algorithm 
\cite{agrawal94fast}, first proposed in the context of discovering association 
rules involving many items in a large data base. This idea has been extended to 
sequential data streams and the frequent episode discovery methods that we propose 
here are based on the same idea. 

We show here that, by adopting such a data mining 
method, we can efficiently {\em discover} important regularities in the multi-neuronal 
spike sequences. We illustrate  this using simulated as well as real
 spike sequences. We use a 
simulator where each neuron is modelled as an inhomogeneous Poisson process whose 
firing rate is modified based on the input received from other neurons. We also 
implement the refractory period by filtering out spikes (generated under the Poisson 
process) that are too close to the previous spike, 
 before they reach any down-stream neuron. Using this simulator, we also show 
that we can assess statistical significance of the detected patterns. (This is 
done using the same idea as in the `jitter' method \cite{date2001}). In addition to 
the results on simulated spike trains, we also show the effectiveness of our approach 
by analyzing some data obtained through micro-electrode array experiments.  

Rest of the paper is organized as follows. Section~\ref{sec:review} presents a brief 
review of analysis of multi-neuronal spike trains. In Section~\ref{sec:epi} we 
briefly explain our method of frequent episode discovery and discuss how this method 
can be used to infer interesting patterns in spike train data. The algorithms for 
discovering frequent episodes under different temporal constraints are explained 
in the next two sections. We explain our 
simulation model and present the results obtained in Section~\ref{sec:simu}. The paper is 
concluded with a discussion of the method and the many possibilities it offers, in 
Section~\ref{sec:dis}.

\section{Multi-neuronal Data Analysis}
\label{sec:review}

Over the last couple of decades, increasingly better methods are becoming available for 
 simultaneously recording the activities of hundreds of neurons 
\cite{Abeles1988,Meister1994,Meister1996,Nicolelis2003,Gerstein2004,WiseAnderson2004,
Ikegaya2004,Hosoya2005,Potter2006}  and hence  development of efficient 
algorithms to analyze multi-neuronal spike trains is becoming critical. This field has a 
long history, beginning with the work of Gerstien and his 
collegues \cite{GersteinPerkel1972} and a recent review \cite{Brown2004} 
summarizes three decades of development in this area. 

Microelectrode array (MEA) is a popular technology for simultaneously recording the 
spike signals from many neurons and has now become a standard method used in 
experiments with neuronal ensembles. 
A typical MEA setup consists of $8\times 8$ grid of 64 electrodes with inter-electrode
spacing of about 25 microns and can be mounted on a neural culture
or brain slice. Other technologies for
recording from multiple neurons include imaging of neuronal currents using
some specialized dyes. One popular method here is to image the Calcium currents. 
 These technologies now allow for gathering of
vast amounts of data, especially in neuronal cultures, using which one wishes
to study connectivity patterns and microcircuits in neural systems 
\cite{Potter2006,Ikegaya2004}.

The availability of vast amounts such data means that developing efficient methods
to analyze neuronal spike trains is a challenging task of immediate utility
in this area \cite{Brown2004}.
A major goal of such neural data analysis is to characterize how neurons
that are part of an ensemble interact with each other.

The patterns that one is interested in  can be roughly grouped into
what are called Synchrony, Order and Synfire chains. Synchronous firing by a group of
neurons is interesting because it can be an efficient way to transmit information
\cite{Meister2005}. 
One can identify synchronous firing of neurons by analyzing cross correlation of 
spike trains \cite{Riehle1997,Riehle2000,Schnitzer2003}. 
Ordered firing sequences of neurons where times between
firing of successive neurons are fairly constant denote a chain of triggering
events and unearthing such relations between neurons can thus reveal some
microcircuits \cite{Abeles2001}.
Discovering temporally ordered firing sequences is important for understanding 
functional connectivity. If neuron A is functionally connected to 
neuron B, it influences the firing of neuron B. If this is an excitatory connection 
(with or without a delay), then, if A fires, B is likely to fire soon after that. Hence, 
discovering the order of neuronal firings can help decipher the functional connectivity. 
Memory traces are probably embedded in such sequential activations of neurons or neuronal 
groups. 
Signals of this form have recently been found in groups of hippocampal neurons by 
Lee and Wilson \cite{Lee2002}. They used specialized algorithms to serch for such 
 ordered firings by a group of neurons (when these orders are known or suspected)
 \cite{Lee2004}.  There are also other algorithms 
for detecting ordered firing sequences with precise timing relationships 
\cite{Abeles1988,Tetko2001}. These methods are based on analyzing cross correlation of 
spike trains where one spike train is delayed with respect to the other. 
An ordered chain of firings of neuronal groups (rather than single neurons) is sometimes 
called a Synfire chain and is believed to be an important microcircuit \cite{Ikegaya2004}. 
A synfire chain can be thought of as a compound pattern involving both synchrony and 
order. 

Discovering such interesting patterns in spike trains 
amounts to unearhing groups of neurons 
that fire in some kind of coordinated fashion. As already mentioned, in most of the 
currently available methods, the curse of dimensionality forces the analysis 
to be confined to a few variables at a time. 
For the same reason, it is often very difficult 
to {\em discover} all patterns of a particular kind. Thus, many of the 
available  algorithms are for 
counting occurrences of specific list of paterns. In the next section we explain the idea 
of frequent episodes and show that this data mining viewpoint 
gives us a unified algorithmic 
scheme for discovering many types of interesting patterns in spike train data.

\section{Frequent Episode Discovery}
\label{sec:epi}

Frequent episode discovery framework was proposed by Mannila et.al. 
\cite{Mannila1997} in the context analyzing alarm sequences in a communication 
network. Laxman et.al. \cite{Srivats2005} introduced the notion of 
non-overlapped occurrences as episode frequency and proposed  
efficient counting algorithms. We first give brief overview of this framework.

In the frequent episodes framework, the data to be analyzed is a sequence of events 
denoted by $\langle(E_{1},t_{1}),(E_{2},t_{2}),\ldots\rangle$ where $E_{i}$ 
represents an \textit{event type} and $t_{i}$ the \textit{time of occurrence} of 
the $i^{th}$ event. $E_i$'s are drawn from a finite set of event types. 
The sequence is ordered with respect to time of occurrences of the events so 
that, $t_i\le t_{i+1}$, for all $i=1,2,\ldots$. The following is an 
example event sequence containing 7 events with 5 event types.
\begin{equation}
\langle(A,1),(B,3),(D,4),(C,6),(A,12),(E,14),(B,15)\rangle
\label{eq:data-seq}
\end{equation}

In multi-neuron data, a spike event has the label of the neuron (or the electrode 
number in case of multi-electrode array recordings) which 
generated the spike as its event type  and has the associated time of occurrence. 
The neurons in the ensemble under observation fire action potentials at 
different times, that is, generate spike events. All these spike events are 
strung together, in time order, to give a single long data sequence as needed for 
frequent episode discovery.

The general temporal patterns that we wish to discover in this 
framework are called episodes. 
 In this paper we shall deal with two types of 
episodes: \textit{Serial} and \textit{Parallel}.

Formally, an episode $\alpha$ is a triple 
$(V_{\alpha},\leq_{\alpha},g_{\alpha})$, where $V_{\alpha}$ is a set of nodes, 
$\leq_{\alpha}$ is a partial order on $V_{\alpha}$, and 
$g_{\alpha}:V_{\alpha}\rightarrow \zeta$ (the set of event types), is a mapping 
associating each node with an event type. 
For an episode to occur in a data stream,  
 the events in $g_{\alpha}(V_{\alpha})$ have to occur in the order described 
by $\leq_{\alpha}$. 
The size of $\alpha$, denoted as $|\alpha|$, is 
$|V_{\alpha}|$ (i.e. the number of nodes in $V_{\alpha}$). Episode $\alpha$ is a 
parallel episode if the partial order $\leq_{\alpha}$ is a null set. It is a 
serial episode if the relation $\leq_{\alpha}$ is a total order. 
A  partial order which is neither a total order nor a null set  
corresponds to the most general class of episodes. Such episodes can be 
described by directed acyclic graphs.

A \textit{serial episode} is an ordered tuple of event types. For example, 
$(A\rightarrow B\rightarrow C)$ is a 3-node serial episode. The arrows in this 
notation indicate the order of the event types. Such an episode is said to 
\textit{occur} in an event sequence if there are  corresponding events in the 
prescribed order. In sequence (\ref{eq:data-seq}), the events 
{${(A,1),(B,3),(C,6)}$} constitute an occurrence of the above episode. In 
contrast a \textit{parallel episode} is similar to an unordered set of items. It 
does not require any specific ordering of the events. We denote a 
3-node parallel episode with event types $A$, $B$ and $C$, as $(ABC)$. An 
occurrence of $(ABC)$ can have the events in any order in the sequence. 

We note here that occurrence of an episode (of either type) does not 
require the associated event types to occur consecutively;  
there can be other intervening events between them. 
 In the multi-neuronal data, if neuron $A$ makes 
neuron $B$ to fire, then, we expect to see $B$ following $A$ often. However, in 
different occurrences of such a substring, there may be different number of 
other spikes between $A$ and $B$ because many other neurons may also be spiking 
simultaneously. Thus, the episode structure allows us to unearth patterns  
in the presence of such noise in spike data.

{\em Subepisode}: 
An episode $\beta$ is a sub-episode of episode $\alpha$ if all event types of 
$\beta$ are in $\alpha$ and if partial order among the event types of $\beta$ is 
same as that for the corresponding event types in $\alpha$. For example 
$(A\rightarrow B)$, $(A\rightarrow C)$, and $(B\rightarrow C)$ are 2-node 
sub-episodes of the 3-node episode $(A\rightarrow B\rightarrow C)$, while 
$(B\rightarrow A)$ is not. In case of parallel episodes, there is no ordering 
requirement. Hence every subset of the set of event types of an episode is a 
subepisode. It is to be noted here that occurrence of an episode implies 
occurrence of all its subepisodes.


{\em Frequency of episodes}: 
A frequent episode is one whose frequency exceeds a user specified threshold. 
The frequency of an episode can be defined in many ways. It is intended
to capture some measure of how often an episode occurs in an event
sequence. One chooses a measure of frequency so that frequent episode discovery is 
computationally efficient and, at the same time, higher frequency would imply that 
an episode is occurring often. 
For the results presented in this paper, we
use the non-overlapped occurences count as the frequency
\cite{Srivats2005,Srivats2006}. 

Two occurrences of an episode are 
said to be \textit{non-overlapped} if no event associated with one appears in 
between the events associated with the other. 
A collection of 
occurrences of $\alpha$ is said to be non-overlapped if every pair of occurrence 
in it is non-overlapped. The corresponding frequency for episode $\alpha$ is 
defined as the cardinality of the largest set of non-overlapped occurrences of 
$\alpha$ in the given event sequence. (See \cite{Srivats2005} for more discussion). 

This definition of frequency results in very efficient 
counting algorithms \cite{Srivats2005}. 
It is also more intuitively satisfying because it counts a 
well defined subset of the set of all  occurrences of an episode.
In the context of our application, counting non-overlapped occurrences is 
natural because we would then be looking at causative chains that happen at 
different times again and again.

\subsection{Temporal Constraints}
As stated earlier, while analyzing neuronal spike data, it is useful to consider 
methods, 
where, while counting the frequency, we include only those occurrences which 
satisfy some additional temporal constraints. We mainly consider two types of 
such constraints: episode expiry time and inter-event time constraints. 

Given an episode occurrence (that is, a set of events in the data stream that 
constitute an occurrence of the episode), we call the largest time difference 
between any two events constituting the occurrence as the span of the occurrence. 
For serial episodes, this would be the difference between times of the 
first and last events of the episode (in an occurrence). 
The episode expiry time constraint 
requires that we count only those occurrences whose span is less than a (user-specified) 
time $T_X$. (In the algorithm in \cite{Mannila1997}, the window width essentially 
implements an upper bound on the span of occurrences.) An efficient algorithm for 
counting non-overlapping occurrences of serial episodes that satisfy an expiry time 
constraint is available in \cite{Srivats2006}. 
However, currently there is no 
algorithm for counting occurrances 
 parallel episodes under expiry time constraint. We present such an 
algorithm in the next section. 

The inter-event time constraint, which is meaningful only 
for serial episodes, is specified by giving an interval of  
the form $(T_{low}, T_{high}]$ and requires that the difference between the times of 
 every pair of successive events in any occurrence of a serial episode 
should be in this interval.  In a generalized form of this constraint, we may have 
different time intervals for different pairs of events. 
In the next section, we present algorithms for 
counting non-overlapped occurrences of episodes with such inter-event time 
constraints. Our algorithm also discovers the most suitable interval constraint 
(choosing from a set of intervals) for each consecutive pair of events in the discovered 
frequent episodes. This leads to discovery of episodes under  generalized 
inter-event time constraints. 

In the next subsection we explain the importance of these temporal constraints 
for capturing many of the desired patterns in spike data in terms of frequent episodes. 
While these temporal constraints are motivated by our application, these are 
fairly general and would be useful in many other applications of frequent 
episode discovery. 

\subsection{Episodes as patterns in neuronal spike data}
 
\begin{figure}[!htb]
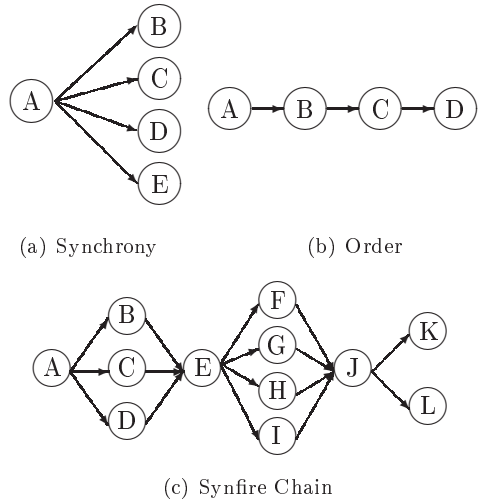

\centering
\subfigure[Synchrony]{\input{parallel-structure.pic}}    
\subfigure[Order]{\input{serial-structure.pic}}
\subfigure[Synfire Chain]{\input{synfire-structure.pic}}
\caption{Examples of neuronal connection structures that can result in different 
patterns in the spike trains 
: (a). simple circuit that can generate synchronous firing patterns. 
Neurons B,C,D,E may fire synchronously, (b). simple 
circuit that generates firing of A, B, C, D in order, (c). A synfire chain pattern where 
different groups of synchronously firing neurons obey a serial order.}
\label{fig:episode-structures}
\end{figure}

The analysis requirements of spike train data
are met very well by the frequent episodes framework. Serial
and parallel episodes with appropriate temporal
constraints can capture many patterns of interest in multi-neuronal data. 
Fig.~\ref{fig:episode-structures} shows some
possibilities of neuronal interconnections that may give rise to
different patterns in spike data.

As stated earlier, one of the patterns of interest is
Synchrony or co-spiking activity in which groups of neurons
fire synchronously. This kind of synchrony may not be precise. 
That is, all neurons in the group need not fire at exactly the same instant of time. 
Allowing for some amount of variability, co-spiking
activity requires that all neurons must fire within a small interval
of time of each other (in any order) for them to be grouped together.
Such synchronous firing patterns may be generated using the structure
as shown in fig.~\ref{fig:episode-structures}(a).
Such patterns of Synchrony can be discovered 
by looking for frequent parallel episodes which satisfy an expiry time constraint.
For example, we can choose the expiry time to be less than a typical
synaptic delay.
The expiry time here controls the amount of variability allowed
for declaring a grouped activity as synchronous.

Another pattern in spike data is ordered firings.
A simple mechanism that can generate ordered firing sequences is
shown in fig.~\ref{fig:episode-structures}(b).
Serial episodes capture such a pattern
very well. Once again, we may need some additional time constraints. 
A useful constraint is that of inter-event time constraint.
In multi-neuron data, if we want to conclude that $A$ is causing $B$ to
fire, then $B$ can not occur too soon after $A$ because there would be
some propagation delay and $B$ can not occur too much later than $A$
because the effect of firing of $A$ would not last indefinitely.
For example, we can prescribe that inter-event times should be in the range
of one to two synaptic delay times so that a frequent serial episode may
capture an underlying microcircuit.
Thus, serial episodes with proper inter-event time constraints can capture
ordered firing sequences which may be due to underlying functional connectivity.

Another important pattern in spiking data is that of synfire
chains \cite{Ikegaya2004}. This
consists of groups of synchronously firing neurons strung together with
tight temporal constraints, repeating often. We can discover such synfire chains 
by combining parallel and serial episode discovery. 

The  structure shown
in Fig.~\ref{fig:episode-structures}(c) captures such a synfire chain. We can
think of this as a microcircuit where $A$ primes synchronous firing of $(B C D)$,
which, through $E$, causes synchronous firing of $(F G H I)$ and so on. When
such a pattern occurs often in the spike train data, parallel episodes like
$(B C D)$ and $(F G H I)$ become frequent (by using appropriate 
expiry time constraint). 
After discovering all such parallel episodes,  we  replace all
recognized occurrences of each of these episodes by a new event in the data 
stream with a new symbol (representing the episode) for the event type and 
an appropriate time of occurrence.  Then we 
discover serial episodes  on this new data stream. With this procedure, 
 we can unearth patterns such as 
synfire chains. 

Summarizing the above discussion, we can assert that
 frequent episode discovery with  various temporal
constraints gives us a lot of flexibility in the kind patterns
that we can discover in multi-neuronal spiking data.

\subsection{Algorithms for frequent episode discovery}

As said earlier, there are efficient frequent episode discovery algorithms
that can handle the required temporal constraints. There are also algorithms that
can discover `useful' inter-event time constraints automatically from
the data. In this subsection, we briefly explain the basic idea in these algorithms. 
We give 
details of the algorithms needed for discovering parallel episodes with 
expiry time constraint and for discovering serial episodes with inter-event time 
constraints in the next section. 
The reader is referred to  \cite{Srivats2005,Deb2006} for more details regarding 
different algorithms for frequent episode discovery.

Consider the problem of discovering all frequent serial episodes upto a given 
size,say, $n$. The discovery process has two main steps. First, we build a set of candidate 
episodes and next we obtain the frequencies (i.e., count the non-overlapping 
occurances) of the candidates in the data 
 so that we can retain only those whose frequencies 
are above the user-set threshold. 

Even if we assume only 50 neurons (or, in the jargon of datamining, event types), 
the number of possible $n$-node serial episodes would be unmanageably large even for 
$n$ as small as 5. As stated 
earlier, it is this combinatorial explosion that limits all the current spike-data analysis 
techniques from being able to discover large sequential patterns. In the datamining 
methods, this is handled by using the idea of discovering progressively larger episodes, 
as we explain below. 

Recall that, given a serial episode, any subsequence which conforms to the 
order of event types in the episode  is called a subepisode. 
For now, let us 
assume that there are no temporal constraints. The key observation is that the episode 
can be frequent only if all its subepisodes are frequent. This is immediately 
obvious because, for example, given two non-overlapping occurances of 
$A \rightarrow B \rightarrow C$, we have atleast two non-overlapping 
occurances of each of its subepisodes. This immediately gives rise to a level-wise 
procedure for discovering all frequent episodes. First we discover all frequent 
1-node episodes. (This is simply a histogram of event types). Then we build a set 
of candidate 2-node episodes such that the 1-node subepisodes of all candidates 
are seen to be frequent. Now through one more pass over the data, we count the 
non-overlapping occurrances of all the candidates and thus come out with frequent 
2-node episodes. Now we combine only the frequent 2-node episodes to build 
a candidate set of 3-node episodes and so on. Thus at stage $n$, using the 
already discovered set of frequent $n$-node episodes, we build the set of 
candidate $(n+1)$-node episodes and by counting their occurrances in the data 
(using one more pass over the data), we come out with frequent $(n+1)$-node 
episodes. This procedure controls the combinatorial explosion because we are, 
after all, interested only in episodes that occur sufficiently often. By choosing 
a suitably large frequency threshold, as the size of episodes grows, the number 
of frequent episodes would come down. (It is highly unlikely that all large 
random sequences occur often in the data). Because of this, the number of 
candidates becomes much much less than the combinatorially possible number, as the 
size of episodes grows. 

Now let us examine whether this idea works even when we impose additional 
temporal constraints. Suppose we use expiry time constraint. Then each occurrance  
of an episode which completes within the expiry time also contains an 
occurrance of its subepisodes each of which also complete within the 
expiry time. Hence, once again all subepisodes would be at least as frequent as 
the episode. Next let us consider the inter-event time constraint. Now, 
it is no longer true that subepisodes are as frequent as episodes. This 
is because we may have many occurrances of $A \rightarrow B \rightarrow C$ where 
each pair of consecutive events occur within time, say, $T_x$, but there may 
be no occurance of the episode $A \rightarrow C$ such that the time between 
the two events is less than $T_x$. So, it may appear that our nice level-wise procedure 
breaks down under inter-event time constraints. However, we observe that 
if we confine ourselves only to prefix and suffix subepisodes then, once again, 
frequency of subepisodes would be atleast as much as that of the episode. 
All we need is a slight change in the candidate generation strategy \cite{Deb2006}.  

In a wide variety of data mining applications this strategy 
is seen to be very effective in controlling 
the combinatorial explosion. This basic idea is from the so called 
Apriori algorithm \cite{agrawal94fast} in the context of dicovering 
frequent itemsets. This is extended to the case of discovering episodes 
by Mannila \cite{Mannila1997}. The idea of non-overlapped occurances 
as frequency makes the process of obtaining frequencies of episodes very 
fast. This and the extensions to tackle various temporal constraints 
are described in \cite{Srivats2005,Srivats2006,Deb2006}. 

\begin{figure}

\begin{centering}\subfigure[Event Sequence]{\label{fig:algo0}\includegraphics[scale=0.75]{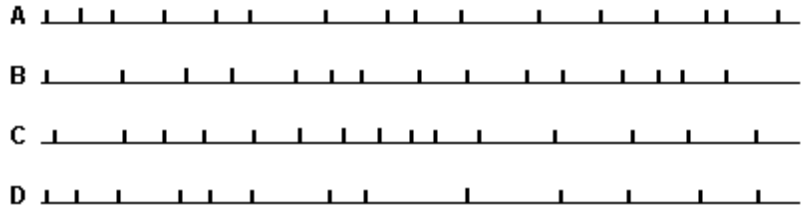}}
\subfigure[Step - 1]{\label{fig:algo1}\includegraphics[scale=0.75]{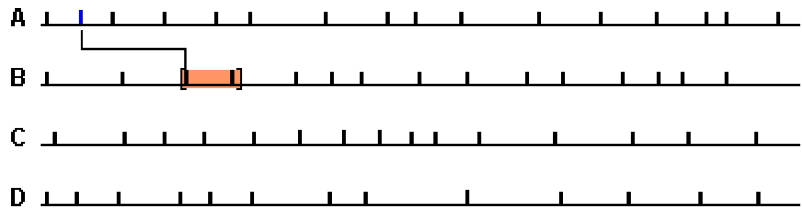}}
\subfigure[Step - 2]{\label{fig:algo2}\includegraphics[scale=0.75]{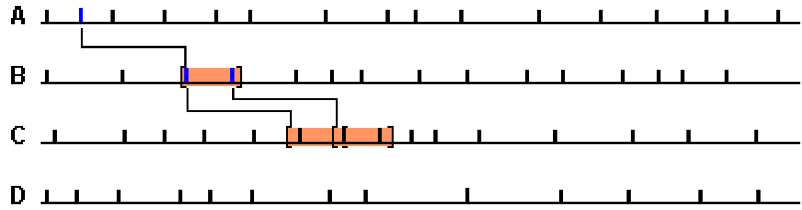}}
\subfigure[Step - 3]{\label{fig:algo3}\includegraphics[scale=0.75]{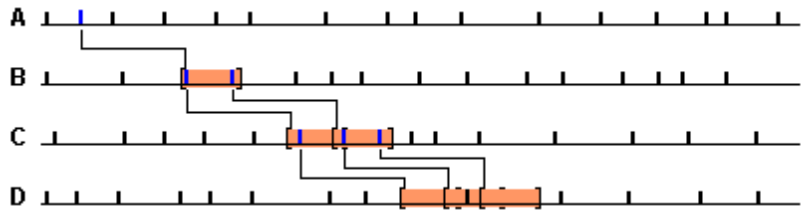}} \par\end{centering}

\caption{Fequent serial Episode discovery - Counting Algorithm. 
Various steps in recognizing an occurrance of a 
serial episode $A\rightarrow B\rightarrow C\rightarrow D$ are shown. After 
seeing $A$, the first episode, the algorithm looks for occurrances of $B$ within 
the time window as specified by the inter-event time constraint. The multiple 
possibilities of $B$ are to be remembered till we find one complete occurrance 
satisfying all inter-event time constraints. In the algorithm, we have to be 
simultaneously looking for such occurrances for a whole set of episodes through 
a single pass over the data. See text for more explanation.}

\label{fig:countalgo}
\end{figure}

Given that we can control the growth of candidates as the size of 
episodes increases, the next question is how do we count the frequencies 
of a set of candidate episodes. This is done by having a finite state 
automaton for each episode such that it recognizes the occurrance of an 
episode. As we traverse the data, for each event (spike) we encounter, we make 
appropriate state changes in all the automata and whenever an automaton 
transits to its end state we increment the count of the corresponding 
episode. Thus, we can simultaneously count the occurrances of a set of candidates 
using a single pass over the data. The number of active automata per episode that we 
need (which is same as the temporary memory needed by the algorithm) depends on what all 
types of occurrances we want to count. Restricting the count to only non-overlapped 
occurrances makes the counting process also very efficient \cite{Srivats2005}. Later, we 
give full details of the candidate generation strategy and the counting procedure 
for two algorithms. Before that, we explain the basic idea of this 
counting (under inter-event 
time constraints) in Fig.~\ref{fig:countalgo}

The first panel in the figure shows the spike data as a raster plot. For illustration, 
consider counting occurrances of the $A \rightarrow B \rightarrow C \rightarrow D$. 
At the beginning, 
there would be an automaton of this episode that is waiting for event type $A$. When 
we see an $A$, we make a state transition and then the automaton is waiting to see a $B$. 
This is shown in panel (b) in the figure. Now, due to the inter-event time 
constraint, this automaton needs a $B$ within some window on the time axis 
as shown in panel (b). When we reach those time points, a $B$ there can now cause 
a state transition in this automaton and it now starts waiting for $C$. However, 
there may be more than one $B$ in the appropriate time window and we have to 
remember all these because, at this stage, we can not know which of these, if any, leads 
to an occurrance of the episode that satisfies the inter-event time constraints. (This 
is the reason we may need more than one active automaton per episode during the 
counting process). 
Logically this means that the automaton spawns multiple 
copies of itself. However, we can design efficient data structures to remember 
only the minimal information.  Now for each of the $B$ events, we can find 
the appropriate time window 
where we need a $C$ to continue. This is shown in panel (c) of the figure. 
Finally, panel (d) of the figure shows how an occurrance of the episode is 
recognized. Note that, once we complete an occurrance, we can forget all the 
extra events we remembered along the way, becuase we are counting only 
non-overlapped occurrances. This gains us a lot of memory efficiency. 

Though the above explanation considered only serial episodes, similar method 
also works for discovery of other types of episodes as well as other types of 
temporal constraints \cite{Srivats2006,Deb2006}.

\section{Algorithms for discovering frequent episodes under temporal constraints}
\label{sec:algos}

In this section we describe our algorithms that discover frequent episodes 
under expiry and inter-event time constraints.\footnote{This section contains 
technical details of the counting algorithms and it is assumed that the reader 
is familiar with such data mining methods. This section can be skipped without 
any loss of continuity.} Since algorithms for 
taking care of expiry time are available in case of serial episodes \cite{Srivats2006}, 
we consider the case of only parallel episodes under expiry time constraint. 
The inter-event time constraints are meaningful only for serial episodes and 
that is the case we consider. Conceptually, all the algorithms essentially use 
finite state automata for recognizing episode occurrences, which is similar 
to the schemes used in \cite{Mannila1997,Srivats2005}. We essentially use the 
same data structures as in those algorithms for keeping track of potential state 
transitions of different automata. As already stated, the method is a two step procedure 
consisting of candidate generation and counting frequencies of a set of candidates. 
This is shown in \textit{Algorithm \ref{alg:Episode-discovery}}.



\begin{algorithm}
\caption{\label{alg:Episode-discovery}Mining Frequent Episodes}
\begin{algorithmic}[1]
\STATE Generate an initial set of (1-node) candidate episodes (N=1)
\REPEAT
\STATE Count the number of occurrences of the set of (N-node) candidate episodes in one pass of the data sequence
\STATE Retain only those episodes whose count is greater than the
frequency threshold and declare them to be frequent episodes
\STATE Using the set of (N-node) frequent episodes, generate the next set
of (N+1-node) candidate episodes
\UNTIL{ There are no candidate episodes remaining}
\STATE Output all the frequent episodes discovered
\end{algorithmic}
\end{algorithm}

\subsection{\label{sec:parallel-with-expiry}Parallel episodes with
expiry}

In this section we present an algorithm that counts the number of non-overlapped occurrences
of a set of parallel episodes in which all the constituting events occur within
time $T_{x}$ of each other. In order to ensure that we count the 
maximum number of occurrences that satisfy the expiry constraint, we need to 
count the inner most occurrences of each episode. 
The algorithm here discovers parallel episodes with non-repeated event types. 
The pseudo-code for the algorithm is listed 
as \textit{Algorithm~\ref{alg:count-parallel-EXPIRY}} in the Appendix.

The algorithm takes as input, the set of candidate episodes, the event sequence 
and the frequency threshold, and outputs the set of frequent episodes. An 
occurrence of a parallel episodes requires all its constituent nodes to appear 
in the event sequence in any order. At any given time, one needs to wait 
for all the nodes of the episode that remain to be seen. 
Thus, in an automaton based algorithm for recognizing occurrences, 
the states of the automaton would denote sets of event types.
In the implementation of the algorithm here, instead of a single automaton waiting for 
for a set of event types, we maintain 
separate entries for each distinct event type of the episode using a $waits(.)$ 
list indexed by event types. For each event type $A$,  
 each entry in the list  $waits(A)$ is of the form $(\alpha, 
count,init)$, where $\alpha$ is an episode waiting for an $A$, ``$count$'' 
takes values 1 or 0 depending on whether an event of this  
type ($A$) has been seen or not, and $init$ indicates the latest time of occurrence of 
this event type.

In \emph{Algorithm}~\ref{alg:count-parallel-EXPIRY}, when an event
type is seen, we update the $init$ field of each entry waiting for it with the 
current time and retain the entries in the $waits$ list. These entries are still 
waiting for their corresponding event types. When we see the same event type 
again, we update the $init$ field of each of the entries in the $waits$ list as 
earlier. This strategy ensures that all the entries for a given parallel episode 
remember only the latest occurrences of their corresponding event types. 
Thus, we effectively track the inner most occurrence.

An occurrence of an episode is complete when there is no entry for an 
episode which has yet to see the first occurrence of its event type and all the 
event times (remembered by $init$ field) occur within $T_{x}$ of each 
other. An episode specific $counter$ is used to keep track of the event 
types already seen. The span of the episode is the difference between the 
smallest and largest $init$ times of the event types for the episode. If the 
span is within the expiry time $T_{x}$, the episode count is incremented and 
all the entries (in the $waits(.)$ lists) for the episode are 
reinitialized.

If the expiry check fails, we cannot reject all the events types of a parallel
occurrence. This is because, in an occurrence of a parallel episode, the
constituent event types can occur in any order in the event stream. Only those
event types which have occurred before $(t_{i}-T_{x})$, should be rejected,
where $t_{i}$ is the time of the latest event type seen by the algorithm.
Effectively, any later occurrence of these events could possibly complete the
parallel episode (without violating the temporal constraint).
When an occurrence of $\alpha$ is complete, $\alpha.freq$ is incremented, 
$\alpha.counter$ is reset and all the entries for the episode are reinitialized.

\subsubsection*{Candidate generation}

The candidate generation scheme is very similar to the one presented in 
\cite{Agrawal95} for itemsets. Let $\alpha$ and $\beta$ be two $k$-node frequent 
episodes having  $(k-1)$ nodes identical. The potential $(k+1)$-node 
candidate is generated by appending to $\alpha$ the $k^{th}$node of $\beta$. 
This new episode is declared as a $(k+1)$-node candidate if all its $k$-node 
subepisodes are already known to be frequent.

\subsection{\label{sec:interval-discovery-algo}Serial Episode with Inter-event 
Constraints}

Under an inter-event time constraint, the time of successive events in any occurrence have to be in a prescribed interval. To take care of this we use a new episodes structure. The episode structure now consists of an ordered set of intervals besides the set of event types. An interval $(t_{low}^i,t_{high}^i]$ is associated with $i^{th}$ pair of consecutive of event types in the episode. For example, a 4-node serial episode is now denoted as follows:
\begin{equation}
(A^{\underrightarrow{(t_{low}^{1},t_{high}^{1}]}}B^{\underrightarrow{(t_{low}^{2},t_{high}^{2}]}}C^{\underrightarrow{(t_{low}^{3},t_{high}^{3}]}}D)
\label{eq:episode-with-interval}
\end{equation}
In a given occurrence of episode $A\rightarrow$ $B\rightarrow$ $C\rightarrow$ $D$ let $t_{A}$, $t_{B}$, $t_{C}$ and $t_{D}$ denote the time of occurrence of corresponding event types. Then this is a valid occurrence of the serial episode with inter-event time constraint given by (\ref{eq:episode-with-interval}), if $t_{low}^{1}$ $<$ $(t_{B}-t_{A})$ $\le$ $t_{high}^{1}$,
$t_{low}^{2}$ $<$ $(t_{C}-t_{B})$ $\le$ $t_{high}^{2}$ and 
$t_{low}^{3}$ $<$ $(t_{D}-t_{C})$ $\le$ $t_{high}^{3}$.

In general, an $N$-node serial episode is associated with, $N-1$ inter-event constraints of the form $(t_{low}^{i},t_{high}^{i}]$. The algorithm we present is for generalized inter-event constraints. The user needs to specify only the granularity of search by providing a set of non-overlapped time intervals to serve as candidate inter-event time intervals. Using a proper candidate generation scheme and the counting algorithm, we can discover frequent episodes along with the set of most appropriate inter-event intervals for each episode. 
This algorithm is easily particularized to the case where inter-event time constraints are 
explicitly specified. 
\subsubsection{Candidate generation scheme}

The generalized inter-event time constraints are a part of the episode structure. In 
the data sequence, if episode $(A^{\underrightarrow{(0,5]}}$  
$B^{\underrightarrow{(5,10]}}$ $C)$ is frequent, the sub-episodes 
$(A^{\underrightarrow{(0,5]}}$ $B)$ and $(B^{\underrightarrow{(5,10]}}$ $C)$ are 
also as frequent, but pairing event type $A$ with $C$ we would get 
$(A^{\underrightarrow{(?,?]}}$ $C)$ as an episode whose inter-event constraints 
are not intuitive. Hence, the Apriori based candidate generation is not suitable 
here.

\begin{figure}[!htb]
\centering
\input{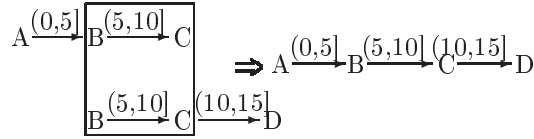}
\caption{\label{fig:Visualization-of-cand-gen-DISCOVERY}Visualization of
Candidate generation for serial episodes with inter-event constraints}
\end{figure}

The candidate episodes in this case are generated as follows. Let
$\alpha$ and $\beta$ be two $k$-node frequent episodes such that
by dropping the first node of $\alpha$ and the last node of $\beta$,
we get exactly the same $(k-1)$-node episode. 
A candidate episode $\gamma$ is generated by copying
the $k$-event types and $(k-1)$-intervals of $\alpha$ into $\gamma$
and then copying the last event type of $\beta$ into the $(k+1)^{th}$
event type of $\gamma$ and the last interval of $\beta$ to the $k^{th}$
interval of $\gamma$. Fig.~\ref{fig:Visualization-of-cand-gen-DISCOVERY} shows 
the candidate generation process graphically.

\subsubsection{Counting episodes with generalized inter-event time constraint}

As already stated, the constraints are
in the form of intervals $(t_{low}^{i},t_{high}^{i}]$, in which the
inter-event times must lie. We first explain the need for a new
algorithm to count occurrences of serial episode with this generalized
structure. Consider the event sequence
\begin{equation}
\langle(A,1),(A,2),(B,4),(A,5),(C,10),(B,12),(C,13),(D,17)\rangle.
\label{eq:interval-discovery}
\end{equation}
Let the serial episode under consideration be $(A^{\underrightarrow{(0,5]}}$ $B^{\underrightarrow{(5,10]}}$ $C^{\underrightarrow{(0,5]}}$ $D)$.
All the current algorithms for counting  occurrences of serial episodes 
either look at left most occurrence of episode or inner most occurrence of 
episode (See \cite{Srivats2006} for details). 
In the given event sequence, the left most occurrence 
is $\langle(A,1),$ $(B,4),$ $(C,10),$ $(D,17)\rangle$ and the inner most 
occurrence  is $\langle(A,5),$ $(B,12),$ $(C,13),$ 
$(D,17)\rangle$, where as the occurrence $\langle(A,2),$ $(B,4),$ $(C,13),$ 
$(D,17)\rangle$ alone satisfies the inter-event interval constraints.

The counting algorithm is listed as Algorithm 2 in the Appendix. 
The  algorithm presented uses $waits$ lists indexed by event types 
 and a linked list of $node$ structures 
for each episode as the basic data-structures.  
The entries in the $waits$ lists are $node$s. For each episode we have a 
doubly inked list of $node$ structures with a $node$ corresponding to each of the event 
types and arranged in the same order as that of the episode. 
The $node$ structure has a $tlist$ field that stores the times of 
occurrence of the event-type represented by its corresponding $node$. 
For example, in the event sequence given by (\ref{eq:interval-discovery}), 
the $node$ representing $A$, after $t=5$, would have $tlist=\{(A,1),(A,2),(A,5)\}$. 
Other field in the $node$ structure is $visited$, 
which is a boolean field that indicates whether the event type is seen atleast once.

\begin{figure}[!htb]
\centering
\psfig{file=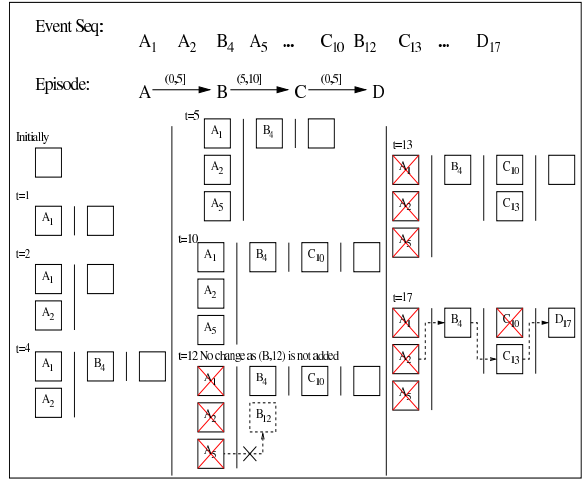, height=2.5in, width=3in}
\caption{\label{fig:Visualization-of-interval-DISCOVERY}Visualization of
Algorithm~\ref{alg:count-serial-INTERVAL-discovery}}
\end{figure}

On seeing an event type $E_{i}$, the algorithm iterates over list $waits(E_{i})$ 
and updates each $node$ in the list. We explain the procedure for updating the $node$s 
by considering the the example sequence given in 
(\ref{eq:interval-discovery}) and the episode 
$\alpha=(A^{\underrightarrow{(0,5]}}$ $B^{\underrightarrow{(5,10]}}$ 
$C^{\underrightarrow{(0,5]}}$ $D)$. Working of the algorithm in this example is 
illustrated in Fig.~\ref{fig:Visualization-of-interval-DISCOVERY}.

The $waits$ lists are initialized by adding the $node$s corresponding
to first event type of each episode in the set of candidates to the corresponding $waits(.)$ list. In the example, let the $node$
tracking event type $A$ be denoted by $node_{A}$, and so on. 
Initially $waits(A)$ contains $node_{A}$.
(That is, the algorithm is waiting for an occurrence of event type
$A$ is the data stream). 
The boxes in 
Fig.~\ref{fig:Visualization-of-interval-DISCOVERY}
represent an entry in the $tlist$ of a $node$. An empty box is one
that is waiting for the first occurrence of an event type. On seeing
$(A,1)$, it is added to $tlist$ of $node_{A}$, and $node_{B}$
is added to $waits(B)$. At any time, \emph{the $node$ structures
are waiting for all event types that have been already seen and the
next unseen event type}.

The algorithm is now waiting for an occurrence of a $B$ and an $A$
as well. At $t=4$, the first occurrence of a $B$ is seen. The $tlist$
of $node_{A}$ is traversed to find atleast one occurrence \emph{of
$A$,} such that $t_{B}-t_{A}\in(0,5]$. Both $(A,1)$ and $(A,2)$
satisfy the inter-event constraint and hence, $(B,4)$ is accepted
into the $node_{B}.tlist$. The rule for accepting an occurrence of
an event type (which is not the first event type of the episode) is
that \emph{there must be atleast one occurrence of the previous event
type} (in this example $A$) \emph{which can be paired with the occurrence
of the current event type} (in this example $B$) \emph{without violating
the inter-event constraint}. Note that this check is not necessary for 
the first event of the episode. 
After seeing the first occurrence of $B$, $node_{C}$ is added to $waits(C)$.
Using the above rules the algorithms accepts $(A,5)$, $(C,10)$ into
the corresponding $tlist$s. At $t=12$, for $(B,12)$ none of the
entries in $node_{A}.tlist$ satisfy the inter-event constraint for
the pair $A\rightarrow B$. Hence $(B,12)$ is not added to the $tlist$
of $node_{B}$. Rest of the steps of the algorithm are illustrated in the figure.

If an occurrence of event type is added to $node.tlist$, it is because there 
exist events for each event type from the first to the event type corresponding 
to the $node$, which satisfy the respective inter-event time constraints.
An occurrence of episode is complete when an occurrence of the last
event type can be added to the $tlist$ of the last $node$ structure
tracking the episode.

The $tlist$ entries shown crossed out in the figure are the ones
that can be deallocated from the memory. 
This is because, given the
inter-event constraint, they can no longer accept an occurrence of
the next event type. 
In the example, at $t=12$, when the algorithm
tries to insert $(B,12)$ into $node_{B}.tlist$, the list of $tlist$
entries for occurrences of $A$'s is traversed. $(A,1)$ with inter-event
constraint $(0,5]$ can no longer be paired with a $B$ since the
inter-event time duration for any incoming event exceeds $5$, hence
$(A,1)$ can be safely removed from the $node_{A}.tlist$. This holds
for $(A,2)$ and $(A,5)$ as well. In this way the algorithm frees
memory wherever possible without additional processing burden.

In order to track episode occurrences we need to store sufficient back references 
in data structures to back track each occurrence. 
This adds some memory overhead, but tracking may be useful in visualizing the discovered episodes.

\section{Results}
\label{sec:simu}

In this section we present some results obtained with our algorithms for analysing 
spike-train data. We present results both on synthetic data generated through 
a simulation model as well as on data gathered from experiments 
on neural cultures. The main 
reason for using  simulator-generated data is that here we can have control 
on the kind of patterns that the data contains and can thus check whether our 
algorithms discover the `true' patterns. The simulation model is intended to 
generated fairly realistic spike trains. For this we actually simulate a network 
of neurons where each neuron is modelled as a Poisson process whose rate changes 
with the input received by the neuron. The network would contain random 
interconnections (which contribute to background spiking) as well as some extra 
strong interconnections among neurons which will contribute to some correlated 
firings by groups of neurons. On simulator-generated data we presents results to show 
that our algorithms can discover different types of embedded patterns. We also present 
some empirical results to argue that the patterns discovered would be statistically 
significant. We then present results on one set of data gathered through Calcium imaging 
techniques and on another set of data gathred through multielectrode array experiments.

\subsection{The spike data generation model}

For the data generation, we use a simulator where each neuron is modelled as an 
inhomogeneous poisson process (whose rate varies with time). In the following paragraphs 
we shall explain the working of our model.

\subsubsection{Simulating Arrivals}
The number of poisson arrivals in time $\Delta t$ is given by
\begin{equation}
P[N(t)-N(t-	\Delta t) = k] = \frac{e^{-\lambda{(t)}\Delta{t}} (\lambda{(t)}\Delta{t})^k}{k!}
	\label{eq:poisson}
\end{equation}

Our simulation is conducted in intervals of $\Delta{t}$. 
Hence the duration of simulation, $T$, is divided into $n$ non-overlapping 
intervals each of size $\Delta{t}$. Let the $i^{th}$ interval (i.e. $[i\Delta{t},(i+1)\Delta{t})$) 
be denoted by $\Delta{t_i}$. It is assumed that the arrival rate remains 
constant over this period (i.e. $\lambda{(t)} = \lambda{_i}, t \in \Delta{t_i}$). 
In a given interval $\Delta{t_i}$, the inter-arrival times are exponentially 
distributed $\approx{exp(\lambda_i)}$. The arrivals in the $i_{th}$ interval 
are simulated as follows.
\begin{equation}
\mbox{Arrivals in the interval }\Delta{t_i}  = \{t_i^1, t_i^2, \cdots, t_i^{K_i}\}, \forall i \in \{1,\dots,n\}
\end{equation}
where each $t_i^j$ is defined as follows
\begin{equation}
t_i^j = t_i^{j-1} + exp(\lambda_{i}), \forall j \in \{1,\cdots,K_i\}
\end{equation} 
and
\begin{equation}
 t_i^{0} = i\Delta{t}, t_i^{K_i} \le (i+1)\Delta{t}
\end{equation}

Hence $K_i$'s are poisson distributed according to equation \ref{eq:poisson}. That is
\begin{equation}
P[K_i = k] = \frac{e^{-\lambda_i\Delta{t}} (\lambda_i\Delta{t})^k}{k!}
\end{equation}

\subsubsection{\label{sec:net-inter-connect}Network Inter-connection}
Our simulation setup consists of a set of $N$ neurons. These neurons are 
inter-connected and a weight is assigned to each inter-connection. Whenever 
a neuron fires, it injects a weighted input into the neurons that it feeds 
into. The inter-connections are setup in such a way that the input from a 
firing neuron will reach a receiver neuron after a certain delay. Each inter-connection 
is capable of having its own delay. Since the simulation is carried on 
in steps of $\Delta{t}$, it makes sense to have these delays as whole number multiples of $\Delta{t}$.

In order to generate noise firings, we randomly interconnect neurons. 
We provide three different schemes for connecting neurons. In the first scheme, 
for a given neuron a number between $0$ and $N$ is randomly chosen. 
Let this number be $k$, then $k$ neurons other than the one in consideration, 
are again randomly picked (with uniform probability) to be the receiver neurons.

In the second scheme, a pair of neurons $(n_i, n_j)$ is picked and with 
probability 0.5 it is decided whether to have a connection from $n_i$ to $n_j$. 
The number inter-connections is, thus, binomially distributed with p
arameters $(N, p = 0.5)$.  The last scheme consists of connecting all pairs 
of neurons and we call this the fully inter-connected network.

In all the three schemes of inter-connection, the weight of a connection 
is a number drawn uniformly from the interval $[-c,c]$.

\subsubsection{Injecting patterns}
When we want to embed any specific pattern, then, we set the weights of the 
required connections between neurons to a higher value. For example, if we 
wish to embed the pattern $A\rightarrow{B}\rightarrow{C}$, we would assign 
higher positive weights to connections $A\rightarrow{B}$ and $B\rightarrow{C}$. 
We shall expain the rational behind the actual weights that we choose in the next section.

\subsubsection{Determining the firing rate of a neuron}

As stated earlier the simulation is carried on in steps of $\Delta{t}$. 
Hence the firing rate of a neuron is determined at the start of each 
$\Delta{t}$ interval and is assumed to remain constant over the interval.  
For random inter-connections the weights are chosen uniformly from the 
interval $[-c,c]$. The weights for causative inter-connections are assigned 
as follows. Let $E_{strong}$ be defined as \textit{the probability of 
firing atleast one spike in $\Delta{t}$ upon receiving one input spike 
from a strong input connection}. The simulator takes $E_{strong}$ as 
in input and determines the weight for a strong connection as follows.

\begin{equation}
	P(N(t + \Delta{t}) - N(t)> 1) = 1 - P(N(t) - N(t - \Delta{t}) = 0)
\end{equation}
\begin{equation}
	E_{strong} = 1 - \frac{e^{-\lambda_{m}\Delta{t}} (\lambda_{m}\Delta{t})^{0}}{0!}
\end{equation}
\begin{equation}
	\lambda_{m} = \frac{-log(1 - E_{strong})}{\Delta{t}}
	\label{eq:lambda_m}
\end{equation}
Here $\lambda_{m}$ is the firing rate required to achieve the desired 
probability of firing atleast one spike in $\Delta{t}$ (i.e. $E_{strong}$).
However due to an absolute refractory period $T_{refractory}$ the number of 
firings in $\Delta{t}$ is actually much less than $\lambda_{m}\Delta{t}$.

Let the firing rate of the $j^{th}$ neuron in $\Delta{t_i}$ intervals be $\lambda_{ji}$. 
This is determined by the following equation.
\begin{equation}
	\lambda_{ji} = \frac{\lambda_{m}}{1 + e^{(-\Delta\lambda{.I_{ji}} + d)}}
	\label{eq:lambda_ji}
\end{equation}

where $I_{ji} = \sum{w_{kj}O_{k}(\widehat{i})}$, $w_{kj}$ is the weight of the 
connection from $k^{th}$ to $j^{th}$ neuron and $O_k(\widehat{i})$ is the number 
of spikes fired by $k^{th}$ neuron in $\Delta{t_{\widehat{i}}}$. Note that 
$(i - \widehat{i})\Delta{t}$ is the delay or time taken for a spike to reach 
neuron $j$ from neuron $k$. Here $d$ is a displacement factor set such that 
with zero input the firing rate of a neuron is $\lambda_{normal}$. And $\Delta\lambda$ 
determines the slope of the sigmoid function. It is currently set to 1.0.

\begin{equation}
	\lambda_{normal} = \frac{\lambda_{m}}{1 + e^{(0 + d)}}
\end{equation}
\begin{equation}
	d = log(\frac{\lambda_{m}}{\lambda_{normal}} - 1)
\end{equation}

Now we set the weight of a strong causative inter-connections such that a single 
input spike achieves a firing rate of $\beta\lambda_{m}$ where $\beta$ is choosen 
close to 1 (i.e. $\approx{0.9}$).

\begin{equation}
	\beta\lambda_{m} = \frac{\lambda_{m}}{1 + e^{(-\Delta\lambda{.w_{strong}.1} + d)}}
\end{equation}
\begin{equation}
	e^{-\Delta\lambda{.w_{strong}}}.e^{d}= \frac{1 - \beta}{\beta}
\end{equation}
\begin{equation}
\end{equation}

Therefore,
\begin{equation}
	w_{strong} = \frac{log(\frac{\beta}{(1 - \beta)}(\frac{\lambda_{m}}{\lambda_{normal}} - 1))}{\Delta{\lambda}}
	\label{eq:w_strong}
\end{equation}

\subsubsection{Adjusting the noise firing rates of neurons in a pattern}

When we embed a pattern by having large-weight interconnection between 
some pairs of neurons, all the neurons that are part of the pattern 
would have their firing rates increased again and again and thus their average firing rate 
would be higher than that of others. This would mean that if we look at the 
histogram of number of spikes by each neuron, we can easily guess which are the 
neurons that participate in the patterned connections. Since our primary 
motivation here is to show the effectiveness of our algorithms in discovering 
hidden patterns, we make a slight modification to the above simulation model 
to make the discovery problem more difficult. We set the normal firing rate 
$\lambda_{normal}$ of all the neurons that are part of a pattern except the 
first neuron according to eq.(\ref{eq:adjust}). 
\begin{equation}
	\lambda_{adjusted} = \alpha{\lambda_{normal}}(1 - E_{strong})
	\label{eq:adjust}
\end{equation}
where, $\alpha(\approx{1.5})$ is a scaling factor.
This rate is achieved by changing the value of $d$ in eq.(\ref{eq:lambda_ji}) 
for the neurons that are part of an embedded pattern. This way, the histogram of 
spikes by different neurons turns out to be almost flat thus giving no 
indication of the embedded patterns. 

\subsubsection{Refactory Period}
In this simulation model an absolute refractory period $T_{refractory}$ is used. 
After a neuron has put out a spike at time $t$, it is not allowed to fire in the 
interval $[t, t+T_{refractory})$. $T_{refractory}$ is usually set to a value close 
to that of $\Delta{t}$.

\subsubsection{Simulated Data}
We use the model to generate data with different patterns as follows.
Let N denote the total number of neurons in the 
system. (We have generated data with N=26, 64 and 100). 
First we randomly interconnect the neurons using one of the schemes 
described in section \ref{sec:net-inter-connect}.
The weight attached to each synapse is set randomly using a 
uniform distribution over $[-c, \; c]$. (We have used $c = 0.50, \; 0.75$). 
When we want to embed any specific pattern, then, we set the weights of the required 
connections between neurons to a higher value. The kind of patterns embedded  
are explained later. 

We set the parameters of the model as follows. The number of neurons, $N$ 
and the range of weights for random interconnections $c$, are 
varied as stated earlier.  
We choose the firing rate of neurons under no input, say, $\lambda_{normal}$. 
(This represents the noise level for the spiking data).
 When we embed a pattern, we want some neurons to cause other neurons 
to fire. This is achieved by increasing their firing rate. For this, we first choose 
a number, $E_{strong} \in [0, \; 1]$, which gives the probability 
that the receiving neuron would 
generate atleast one spike in the next $\Delta{t}$ interval if it receives the expected 
pattern input. The value of $E_{strong}$ then determines the firing rate $\lambda_m$ that 
the neuron should have (by using the Poisson distribution). 
We then determine the weight of connection needed so that 
if each of the intended input neurons sends out one spike (in the appropriate 
time interval) then the receiving neuron 
would reach close to the firing rate of $\lambda_m$ under our chosen sigmoidal function.

For the simulations discused here, we used the following values for parameters:
 $\lambda_{normal} = 20 Hz$, $E_{strong} = 0.95$. (Recall that $E_{strong}$ determines $\lambda_m$ which 
in turn determines weights for the patterned connections). We have chosen 
$\Delta T = 1$ milli sec and chosen the refractory period ($T_{refractory}$) also the same. 
(This would mean that in any $\Delta{t}$ interval there would be atmost one spike from any 
neuron). We have chosen inter-connection delay $5\Delta{t}$ which implies a synaptic delay of 
5 milli sec.

The patterns we want to embed are the kind shown in Fig.~\ref{fig:episode-structures}.
These are realizable by essentially two types of pattern dependent 
interconnections between neurons. One 
is where a neuron primes one (as in a serial episode) or many (as in a parallel 
episode) neurons. Here the weight is determined by requiring that one spike (in the 
appropriate interval) by the priming neuron would increase the firing rate of the 
receiving neuron to $\lambda_m$ so that in the next $\Delta{t}$ interval the 
receiving neuron spikes atleast once with probability $E_{strong}$. The other kind of 
interconnection is where many neurons together prime one neuron (which is used 
in Synfire chains). Here, the weight of each connection is set in such a way that 
only if each of the input neurons spikes once in the appropriate interval then the 
firing rate of the receiving neurons would go upto $\lambda_m$. (If only a few of 
the input neurons fire, then the firing rate of the receiving neuron goes up but 
not all the way upto $\lambda_m$). 

The weights of random connections are set using a mean-zero distribution and hence, 
in an expected sense all neurons keep firing at the `noise' rate of $\lambda_{normal}$. However, 
since the actual input can still assume small positive and negative values, this 
background firing rate would also be fluctuating around $\lambda_{normal}$. Since all firings 
are stochastic, even when a pattern is embedded, the entire patterned firing sequence 
will not always happen. Also, within a pattern of firing of neurons (as per the embedded 
pattern), there would be other neurons that would be spiking randomly. {\em Also, due to 
our implementing of refractory period, the actual firings of neurons are not Poisson}.

\subsection{Discovering patterns in the simulated spike trains}

In this section we present some results to illustrate the effectiveness of 
our datamining algorithms in discovering patterns in spike data. We show that a 
combination of parallel and serial episodes with appropriate temporal 
constraints can capture most of the interesting patterns in spike data. 
We used the 
simulator described earlier to generate the data. The types of interconnections 
among neurons that we used to generate data with different embedded patterns 
are shown in fig.~\ref{fig:episode-structures}. 

 As explained earlier, synchronous firing patterns are well described by 
parallel episodes. To embed a synchrony pattern 
in the spike data, we use the interconnection scheme given in 
fig.~\ref{fig:episode-structures}(a). Here neuron $A$ has strong connections into 
neurons $B,C,D$. Thus, a spike from $A$ would cause, after a synaptic delay, the 
other neurons to spike. Because of the way we choose weights for such pattern-based 
interconnections, this means that with a high probability $B$, $C$ and $D$ would all 
fire within one $\Delta{t}$ (which is 1 milli sec here) interval. Hence we can 
discover such patterns by using the method for discovering parallel episodes with an 
expiry time of less than 1 milli sec. We refer to the circuit shown in the figure as a 
parallel episode of size 3 (because it involves synchronous firing of three neurons).
We can similarly create larger patterns of synchrony.

To create spike data with ordered firing  patterns, we use the interconnection 
scheme similar to the one shown in fig.~\ref{fig:episode-structures}(b). Here, a series of 
neurons are connected through high weights. As explained earlier, serial episodes 
capture such patterns. Hence, to discover patterns of ordered firings, we use 
algorithm for discovering frequent serial episodes with inter-event time constraints. 
Since we have chosen synaptic delay to be 5 milli sec and use 1 milli sec windows for 
gathering input into neurons, we can typically 
use an inter-event interval of 4 -- 6 milli sec 
as the constraint. 

 To create data with synfire 
chains, we use the interconnection scheme as illustrated in 
fig.~\ref{fig:episode-structures}(c). Here, $(B C D)$, $(F G H I)$ etc are 
synchronous patterns that are strung together in a tight temporal order. 
The fixing of weights in this connection scheme is as explained in the previous 
subsection. Essentially, firing of $A$ would send strong inputs into each of 
$B, C, D$. With a high probability they fire synchronously, that is within a 
window of 1 milli sec. The weghts from $B, C, D$ to $E$ are such that only if, 
in fact, they fire synchronously then with a high probability $E$ would fire 
thus triggering the next synchrony. We also note here that $E$ would fire 
within 4 -- 6 mili sec of the synchronous firing of $(B, C, D)$ because there 
is a synaptic delay involved here. 

We can discover such patterns as follows. 
We first discover all (frequent) parallel episodes with expiry time of 1 milli sec or less. 
This would capture the synchrony patterns. Then for each such parallel episode, 
we take each of the occurrances of the episode counted by our algorithm and replace 
all these events (spikes) with a single event with a new name whose time of occurrance 
is put as the midpoint of the corresponding occurrance span of the parallel episode. 
Now on this modified data stream we discover serial episodes with inter-event 
time constraint of 4 -- 6 milli sec. Such a procedure can discover patterns in the 
form of Synfire chains.



We first show that our method can discover specific network patterns that are 
embedded in the data generation process. We also illustrate the ability of our method to 
automatically discover inter-event time constraints most appropriate given the data. 
We discuss three examples for this.


\subsubsection*{Example 1}
\begin{figure}[!htb]
\centering
\input{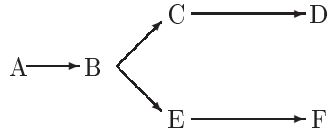}
\caption{\label{fig:pattern-1}Network pattern for Example 1}
\end{figure}

In a 26 neurons network (where each neuron corresponds to an alphabet)
we embed the pattern shown in Fig.\ref{fig:pattern-1}. The simulation
is run for 50 sec and approximately 25,000 spikes are generated. The
synaptic delay is set to be about 5 milli sec. We have chosen 
$\Delta{t} = 1$ milli sec and have taken refractory time also to be 
the same.

\begin{table}[!htb]
\centering
\begin{tabular}{|r|r|r|r|l|}
 \hline
Episode & Freq. & Time & Size & Patterns \\
expiry & Th. & (sec) & (No.) & Discovered \\
\hline
0.0001 & 0.01 & 0.23 & 1(26) & \small{no episode of 2} \\
 & & & & \small{or more nodes} \\ \hline
0.001 & 0.01 & 0.29 & 2(2) & \small{E C : 799; F D : 624} \\ \hline
0.002 & 0.01 & 0.28 & 2(2) & \small{E C : 804; F D : 643} \\ \hline
0.007 & 0.01 & 0.37 & 2(2) & \small{F E D C : 615} \\ \hline
\end{tabular}  
\caption{\label{tab:pattern-1-parallel}Parallel episodes dicovered with 
different expiry time constraints in  Example 1}
\end{table}

\begin{table}[!htb]
\centering
\begin{tabular}{|r|r|r|r|l|}
 \hline
Inter-event & Freq. & Time & Size & Patterns \\
interval & Th. & (sec) & (No.) & Discovered \\
 \hline
0.000-0.001 & 0.01 & 0.29 & 2(4) & \small{C E : 410; E C : 400} \\
 &  &  &  & \small{D F : 329; F D : 303} \\ \hline
0.000-0.002 & 0.01 & 0.31 & 2(4) & \small{C E : 422; E C : 408} \\
 &  &  &  & \small{D F : 348; F D : 323} \\ \hline
0.002-0.004 & 0.01 & 0.26 & 1(26) & \small{no 2 or more} \\ 
	&	&	&	& \small{node episodes} \\ \hline
0.004-0.006 & 0.01 & 0.29 & 4(4) & \small{A B C D : 597} \\
 &  &  &  & \small{A B E F : 589} \\
 &  &  &  & \small{A B E D : 530} \\
 &  &  &  & \small{A B C F : 530} \\ \hline
\end{tabular}  
\caption{\label{tab:pattern-1-serial}Serial episodes discovered with 
different inter-event constraints in Example 1}
\end{table}

The sequence is then mined for frequent parallel episodes with
different expiry times. The results are given in Table
\ref{tab:pattern-1-parallel}. The table shows the expiry time used, 
the frequency threshold, time taken by the algorithm on a Intel dual core 
PC running at 1.6 GHz, the size of the largest frequent episode discovered and 
the number of episodes of this size along with the actual episodes. We follow 
the same structure for all the tables in the three examples. 
The frequency threshold is expressed as
a fraction of the entire data length. A threshold of 0.01 over a data
length of 25,000 spike events requires an episode to occur atleast 250
times before it is declared as frequent. From Table
\ref{tab:pattern-1-parallel} it can be seen that $(CE)$ and $(DF)$
turn out to be the only frequent parallel episodes if the 
expiry time is 1 to 2 milli sec. If the expiry time is too small, we get 
no frequent episodes (at this threshold). On the other hand, if we increase 
the expiry time to be 7 milli sec which is greater than a synaptic delay, 
then even $(FEDC)$ turns out to be a parallel episode. This shows that by using 
appropriate expiry time, parallel episodes discovered 
capture synchronous firing patterns. 

The results of serial episode discovery are shown in Table
\ref{tab:pattern-1-serial}. With an inter-event constraint of 4-6
milli sec, we discover all paths in the network (Fig.
\ref{fig:pattern-1}). When we prescribe that inter-event time be less 
than 2 milli sec (when synaptic delay is 5 milli sec), we get nodes in the 
same level as our serial episodes. If we use intervals of 2-4 milli sec, we 
get no episodes because synchronous firings mostly occur much closer 
and firings related by a synapse have a delay of 5 milli sec. Thus, using 
inter-event time constraints, we can get fair amount of information of the 
underlying connection structure.  
It may seem surprising that we also discover
$A\rightarrow B$ $\rightarrow C$ $\rightarrow F$ and $A\rightarrow B$
$\rightarrow E$ $\rightarrow D$ when we use 4--6 milli sec constraint. 
This is because, the network structure is 
such that $D$ and $F$ fire about one synaptic delay time 
after the firing of $C$ and $E$. Thus, the serial episodes give the sequential 
structure in the firings which could, of course, be generated by different 
interconnections. The frequent episodes discovered 
provide a handle to unearthing the hierarchy seen in the data (i.e.
which events co-occur and which ones follow one another).

\subsubsection*{Example 2}
In this example we consider the network connectivity pattern as shown in
Fig.~\ref{fig:episode-structures}(c). As stated earlier, this is an example 
of possible network connectivity that can generate
Synfire chains. We use the same parameters in the simulator as in Example 1
 and generate spike trains data using this connectivity pattern.  
Table~\ref{tab:pattern-3-parallel} shows the parallel episodes discovered  
 and Table~\ref{tab:pattern-3-serial} shows the serial episodes discovered 
with different inter-event constraints. From the tables, it is easily seen 
that parallel episodes with expiry time of 1 milli sec and 
serial episodes with inter-event time constraint of about one synaptic delay, 
together give good information about underlying network structure. In this example, 
we illustrate how our algorithms can discover synfire chain patterns. As 
explained earlier, we first 
discover all parallel episodes with expiry time 1 milli sec. Then for  
each frequent parallel episode,  
we replace each of its occurrences in the data stream by a new event with event type 
being the name of the parallel episode. This new event is put in with a time 
of occurrence which is the mean time in the episode occurrence. We then discover 
all serial episodes with different inter-event time constraints. The 
results obtained with this method are shown in Table~\ref{tab:pattern-3-synfire}. 
As can be seen, 
the only pattern we discover is the underlying synfire chain. This example shows 
that by proper combination of parallel and serial episodes, we can obtain fairly 
rich pattern structures which are of interest in neuronal spike train analysis. 

\begin{table}[!htb]
\centering
\begin{tabular}{|r|r|r|r|l|}
 \hline
Episode & Freq. & Time & Size & Patterns \\
expiry & Th. & (sec) & (No.) & Discovered \\
\hline
0.001 & 0.01 & 0.15 & 4(1) & \small{L K : 307} \\
 &  &  &  & \small{C B D : 293} \\
 &  &  &  & \small{H G F I : 268} \\
 &  &  &  & \small{rest are} \\
 &  &  &  & \small{sub-episodes} \\ \hline
\end{tabular}  
\caption{\label{tab:pattern-3-parallel}Parallel episodes discovered in Example 2}
\end{table}

\begin{table}[!htb]
\centering
\begin{tabular}{|r|r|r|r|l|}
 \hline
Inter-event & Freq. & Time & Size & Patterns \\
interval & Th. & (sec) & (No.) & Discovered \\
\hline
0.002-0.004 & 0.01 & 0.157 & 1(26) & \small{no episodes of 2} \\
 & & & & \small{or more nodes} \\ \hline
0.004-0.006 & 0.01 & 0.469 & 6(24) & \small{A D E H J K : 195} \\
 &  &  &  & \small{A D E I J K : 194} \\
 &  &  &  & \small{A D E H J L : 193} \\
 &  &  &  & \small{A C E H J K : 192} \\ \hline
0.006-0.008 & 0.01 & 0.156 & 1(26) & no episodes of 2 \\
 & & & & or more nodes \\ \hline
\end{tabular}  
\caption{\label{tab:pattern-3-serial}Serial episodes discovered under 
different inter-event constraints in  Example 2}
\end{table}

\begin{table}[!htb]
\centering
\begin{tabular}{|r|r|r|r|l|}
 \hline
Inter-event & Freq. & Time & Size & Patterns \\
interval & Th. & (sec) & (No.) & Discovered \\
\hline
0.002-0.004 & 0.01 & 0.11 & 1(20) & \small{no episodes of} \\
 & & & & \small{2 or more nodes} \\ \hline
0.004-0.006 & 0.01 & 0.14 & 6(1) & \small{A [C B D] E} \\
 & & & & \small{[H G F I] J [L K] : 137} \\ \hline
0.006-0.008 & 0.01 & 0.12 & 1(20) & \small{no episodes of} \\
 & & & & \small{2 or more nodes} \\ \hline
\end{tabular}  
\caption{\label{tab:pattern-3-synfire}Synfire chain episodes discovered in  Example 2}
\end{table}

\subsubsection*{Example 3}
\begin{figure}[!htb]
\centering
\input{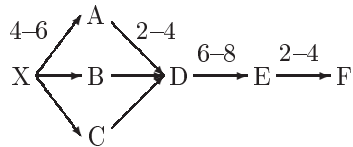}
\caption{\label{fig:pattern-2}Network Pattern for Example 3}
\end{figure}

In this example, we choose a network pattern where different pairs of 
interconnected neurons can have different synaptic delays 
and we demonstrate the ability of our algorithm to 
automatically discover appropriate inter-event intervals. 
The pattern is shown in Fig.
\ref{fig:pattern-2}, where we have different synaptic delays 
as indicated on the figure, for different inter-connections.

\begin{table}[!htb]
\centering
\begin{tabular}{|r|r|r|r|l|}
 \hline
Episode & Freq. & Time & Size & Patterns \\
expiry & Th. & (sec) & (No.) & Discovered \\
\hline
0.001 & 0.01 & 0.28 & 3(1) & \small{A B C : 614} \\ \hline
0.002 & 0.01 & 0.25 & 3(1) & \small{A B C : 617} \\ \hline
0.004 & 0.01 & 0.28 & 4(1) & \small{A B C D : 537} \\ \hline
0.006 & 0.01 & 0.32 & 4(2) & \small{X A B C : 602} \\
 &  &  &  & A B C D : 542 \\ \hline
\end{tabular}  
\caption{Parallel episodes discovered under different expiry times in  Example 3}
\label{tab:pattern-2-parallel}
\end{table}

\begin{table}[!htb]
\centering
\begin{tabular}{|r|r|r|r|l|}
 \hline
Inter-event & Freq. & Time & Size & Patterns \\
interval & Th. & (sec) & (No.) & Discovered \\
\hline
0.000-0.002 & 0.01 & 0.32 & 2(6) & \small{A C : 385; B A : 376} \\
 &  &  &  & \small{B C : 373; A B : 372} \\
 &  &  &  & \small{C A : 361; C B : 355} \\ \hline
0.002-0.004 & 0.01 & 0.37 & 2(4) & \small{E F : 783; A D : 656} \\
 &  &  &  & \small{C D : 651; B D : 646} \\ \hline
0.004-0.006 & 0.01 & 0.28 & 2(3) & \small{X A : 790; X B : 774} \\
 &  &  &  & \small{X C : 769} \\ \hline
0.006-0.008 & 0.01 & 0.29 & 2(2) & \small{D E : 720; X D : 454} \\ \hline
\end{tabular}  
\caption{Serial Episodes discovered under different inter-event time 
constraints in Example 3}
\label{tab:pattern-2-serial}
\end{table}

\begin{table}[!htb]
\centering
\begin{tabular}{|r|r|r|r|l|}
 \hline
Inter-event & Freq. & Time & Size & Patterns \\
interval & Th. & (sec) & (No.) & Discovered \\
\hline
\{0.000-0.002, & 0.01 & 1.37 & 5(1) &  \\
0.002-0.004, &  &  &  & \small{$X^{\underrightarrow{0.004-0.006}} [A B C]$}  \\
0.004-0.006, &  &  &  & \small{$^{\underrightarrow{0.002-0.004}} D^{\underrightarrow{0.006-0.008}}$} \\
0.006-0.008, &  &  &  & \small{$E^{\underrightarrow{0.002-0.004}} F$ : 372} \\
0.008-0.010\} &  &  &  &  \\ \hline
\end{tabular}  
\caption{\label{tab:pattern-2-synfire}Synfire chain episodes discovered in  Example 3}
\end{table}

The results for parallel episode discovery (see Table
\ref{tab:pattern-2-parallel}) show that $(ABC)$ is the group of
neurons that co-spike together. The serial episode discovery results
are given in Table \ref{tab:pattern-2-serial}. As can be seen from 
the table, with different 
 pre-specified inter-event time constraints we can discover only 
different parts of the underlying network graph because no single 
inter-event constraint captures the full pattern. 

As in Example 2, we replace occurrences of parallel episode with a 
new event in the data stream. We then run Algorithm2 to 
discover serial episodes along with inter-event constraints, given   
 a set of possible inter-event intervals. 
The results obtained are shown in 
Table \ref{tab:pattern-2-synfire}. As can be seen from the table, 
the algorithm is very effective in unearthing the underlying 
network pattern.

\subsubsection{Performance of algorithms with multiple patterns}

The earlier examples clearly demonstrate the ability of our data mining 
algorithms in unearthing the connectivity pattern in the neuronal network 
that generated the data. To keep the examples simple we considered only 
single patterns and on single data sets.  Next we 
demonstrate the performance of the algorithm, averaged over many independently generated 
random datasets, when multiple patterns of different 
sizes are present. We illustrate this for all the three types of patterns. 
We  use network with 64 neurons here since most typical micro electrode arrays 
have 64 channels. 
 
First we consider a 64 neuron system with one or more circuits of synchrony  
 embedded. We have varied the size of the synchrony pattern from 8 to 12 and 
have experimented with embedding upto four distinct patterns. For each pattern 
to be embedded, the actual neurons that participate in the pattern are chosen 
randomly. With such circuits in place we generated many data sets with each data 
set of 50 sec duration. (Since the normal firing rate is 20 Hz, in 50 sec each neuron
would, on the average, spike 1000 times thus giving us a data set of about 60,000 
spikes, which is the typical size of spike data sets analyzed). 

The results of our parallel episode discovery are shown in Table~\ref{tab:parallel}.
As explained earlier, our algorithm systematically discovers parallel episodes of 
all sizes whose frequencies are above the threshold set. We have chosen a threshold 
of 300. (We discuss choosing of the threshold in the next subsection). The first four 
columns of the table are self-explanatory. The last column shows the percentage of 
the discovered frequent episodes that are part of the embedded pattern for 
various sizes. As can be seen from the table, even at size three, all episodes discovered 
are part of the embedded pattern. Thus, all long synchrony patterns we discover are 
all `correct' in the sense that they are actually present in the neural system 
that generated the spike data. 

We also like to point out that the time taken by our method to discover all the parallel 
episodes is only about 30 sec on a PC even for discovering four different synchrony 
patterns each involving 12 neurons. This illustrates the fact that these algorithms 
are very efficient in unearthing the patterns. 

\begin{table}
\begin{tabular}{|r|r|r|r|>{\raggedleft}p{1.2cm}|>{\raggedleft}p{1.2cm}|>{\raggedleft}p{1.6cm}|}
\hline 
\multicolumn{1}{|p{1.5cm}|}{Pattern Type}&
\multicolumn{1}{p{1cm}|}{Size}&
\multicolumn{1}{p{1.5cm}|}{No. of distinct patterns}&
\multicolumn{1}{p{1cm}|}{Time taken (in sec)}&
\multicolumn{3}{p{4cm}|}{Fraction of detected frequent episodes that are part of the embedded
pattern}\tabularnewline
\hline
\multicolumn{1}{|c|}{}&
\multicolumn{1}{c|}{}&
\multicolumn{1}{c|}{}&
\multicolumn{1}{c|}{}&
Pattern Length &
Min Count &
\% Fraction \tabularnewline
\cline{5-5} \cline{6-6} \cline{7-7} 
Parallel &
8 &
2 &
1.968 &
1 &
941 &
25.0\%\tabularnewline
&
&
&
&
2 &
785 &
100.0\%\tabularnewline
&
&
&
&
8 &
520 &
100.0\%\tabularnewline
&
&
3 &
2.438 &
1 &
1001 &
37.5\%\tabularnewline
&
&
&
&
2 &
841 &
100.0\%\tabularnewline
&
&
&
&
8 &
542 &
100.0\%\tabularnewline
&
&
4 &
2.844 &
1 &
963 &
50.0\%\tabularnewline
&
&
&
&
2 &
812 &
100.0\%\tabularnewline
&
&
&
&
8 &
550 &
100.0\%\tabularnewline
\hline 
Parallel &
10 &
2 &
5.156 &
1 &
987 &
31.2\%\tabularnewline
&
&
&
&
2 &
829 &
100.0\%\tabularnewline
&
&
&
&
10 &
507 &
100.0\%\tabularnewline
&
&
3 &
7.141 &
1 &
985 &
46.9\%\tabularnewline
&
&
&
&
2 &
829 &
100.0\%\tabularnewline
&
&
&
&
10 &
480 &
100.0\%\tabularnewline
&
&
4 &
9.219 &
1 &
970 &
62.5\%\tabularnewline
&
&
&
&
2 &
823 &
100.0\%\tabularnewline
&
&
&
&
10 &
465 &
100.0\%\tabularnewline
\hline 
Parallel &
12 &
2 &
18.64 &
1 &
906 &
37.5\%\tabularnewline
&
&
&
&
2 &
765 &
100.0\%\tabularnewline
&
&
&
&
12 &
408 &
100.0\%\tabularnewline
&
&
3 &
27.875 &
1 &
963 &
56.2\%\tabularnewline
&
&
&
&
2 &
802 &
100.0\%\tabularnewline
&
&
&
&
12 &
400 &
100.0\%\tabularnewline
&
&
4 &
37.578 &
1 &
929 &
75.0\%\tabularnewline
&
&
&
&
2 &
785 &
100.0\%\tabularnewline
&
&
&
&
12 &
389 &
100.0\%\tabularnewline
\hline
\end{tabular}

\caption{Discovery of Synchronous firing patterns: Averaged results over 100 
datasets for discovering parallel episodes of different sizes with multiple patterns 
embedded. The first three columns show type of pattern, number of distinct 
patterns embedded and  time taken on one dataset. When we embed a pattern of 
size, say, 8, the algorithm would discover parallel episodes at all sizes 
upto 8 in its level-wise iterations. The last column shows the minimum frequency 
of discovered patterns for these smaller sizes and also the fraction of discovered 
episodes of that size which are part of the embedded pattern.  
 }
\label{tab:parallel}
\end{table}

\begin{figure}
\subfigure[]{\includegraphics[scale=0.75]{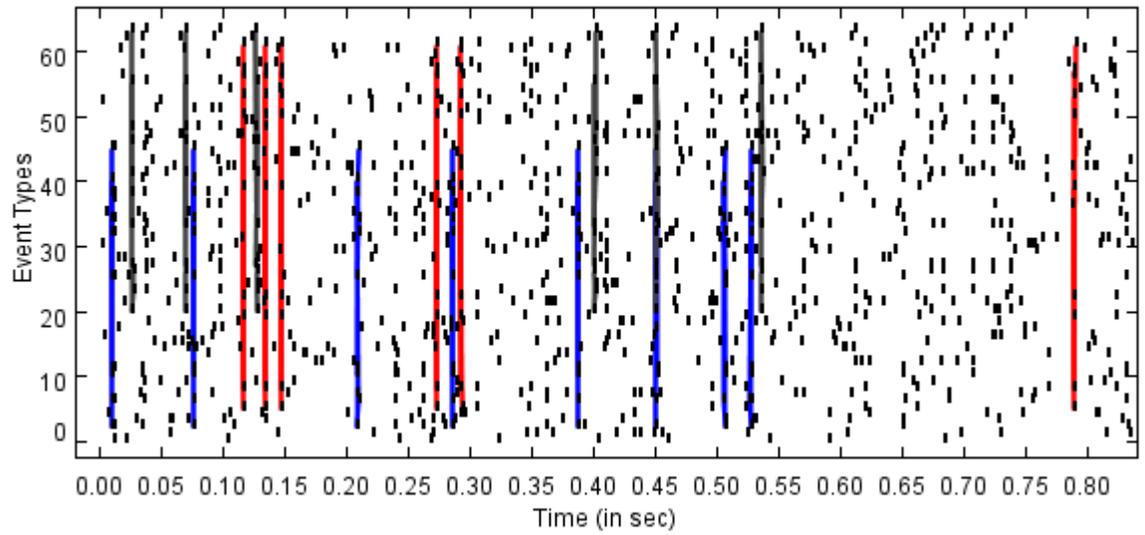}}

\subfigure[]{\includegraphics[scale=0.75]{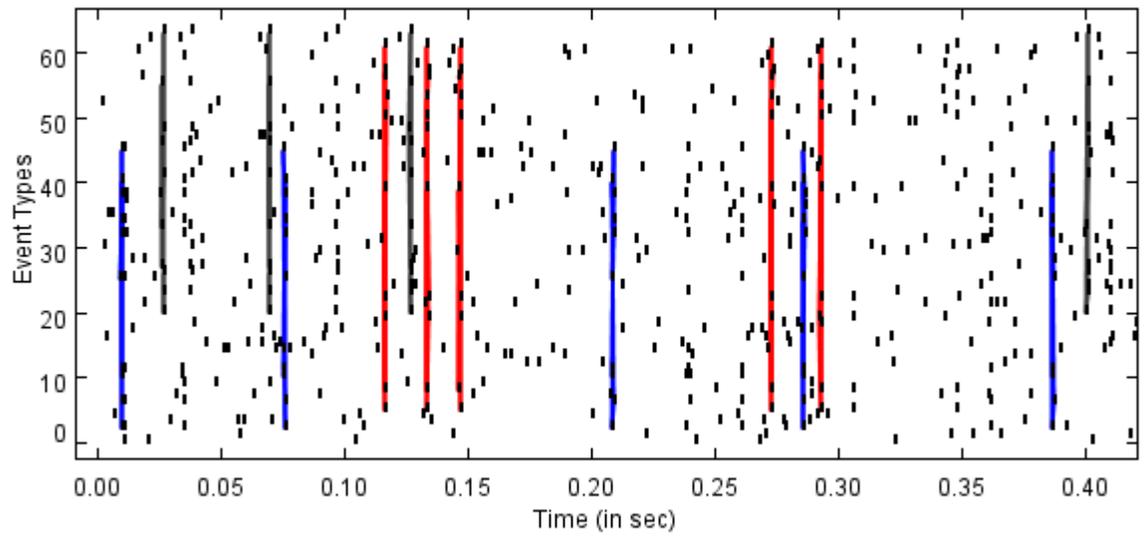}}

\caption{Some of the occurrances of the synchronous firing patterns in a 
typical dataset. }
\label{fig:synchrony-raster}
\end{figure}

In fig.~\ref{fig:synchrony-raster} we show some of the occurrences 
of three different synchronous firing patterns as a raster plot. We 
show them in two different windows on the time axis. 

We next illustrate discovering ordered firing sequences with tight temporal 
relationships. 
We embed many such serial patterns of different 
sizes to test our method. These results are shown in Table~\ref{tab:serial}. 
Once again, from the results it is easily seen that the method is very effective 
in unearthing the patterns of interest. Also, the time taken here is much smaller (less 
than 3 sec). This is because, the tight inter-event time constraints control 
the growth of candidate patterns in the frequent episode discovery method.  

\begin{table}
\begin{tabular}{|r|r|r|r|>{\raggedleft}p{1.2cm}|>{\raggedleft}p{1.2cm}|>{\raggedleft}p{1.6cm}|}
\hline 
\multicolumn{1}{|p{1.5cm}|}{Pattern Type}&
\multicolumn{1}{p{1cm}|}{Size}&
\multicolumn{1}{p{1.5cm}|}{No. of distinct patterns}&
\multicolumn{1}{p{1cm}|}{Time taken (in sec)}&
\multicolumn{3}{p{4cm}|}{Fraction of detected frequent episodes that are part of the embedded
pattern}\tabularnewline
\cline{1-1} \cline{2-2} \cline{4-4} \cline{5-7} 
\cline{3-3} \cline{5-5} \cline{6-6} \cline{7-7} 
\multicolumn{1}{|c|}{}&
\multicolumn{1}{c|}{}&
\multicolumn{1}{c|}{}&
\multicolumn{1}{|c|}{}&
Pattern Length &
Min Count &
\% Fraction \tabularnewline
\cline{5-5} \cline{6-6} \cline{7-7} 
Serial &
8 &
2 &
1.766 &
1 &
1001 &
25.0\% \tabularnewline
&
&
&
&
2 &
867 &
100.0\% \tabularnewline
&
&
&
&
8 &
426 &
100.0\% \tabularnewline
&
&
3 &
1.797 &
1 &
1000 &
37.5\% \tabularnewline
&
&
&
&
2 &
879 &
100.0\% \tabularnewline
&
&
&
&
8 &
443 &
100.0\% \tabularnewline
&
&
4 &
1.891 &
1 &
957 &
50.0\% \tabularnewline
&
&
&
&
2 &
851 &
100.0\% \tabularnewline
&
&
&
&
8 &
430 &
100.0\% \tabularnewline
\hline 
Serial &
10 &
2 &
1.844 &
1 &
984 &
31.2\% \tabularnewline
&
&
&
&
2 &
866 &
100.0\% \tabularnewline
&
&
&
&
10 &
377 &
100.0\% \tabularnewline
&
&
3 &
1.953 &
1 &
935 &
46.9\% \tabularnewline
&
&
&
&
2 &
817 &
100.0\% \tabularnewline
&
&
&
&
10 &
358 &
100.0\% \tabularnewline
&
&
4 &
2.031 &
1 &
945 &
62.5\% \tabularnewline
&
&
&
&
2 &
817 &
100.0\% \tabularnewline
&
&
&
&
10 &
345 &
100.0\% \tabularnewline
\hline 
Serial &
12 &
2 &
1.922 &
1 &
1007 &
37.5\% \tabularnewline
&
&
&
&
2 &
884 &
100.0\% \tabularnewline
&
&
&
&
12 &
311 &
100.0\% \tabularnewline
&
&
3 &
2.203 &
1 &
976 &
56.2\% \tabularnewline
&
&
&
&
2 &
853 &
100.0\% \tabularnewline
&
&
&
&
12 &
309 &
100.0\% \tabularnewline
&
&
4 &
2.297 &
1 &
967 &
75.0\% \tabularnewline
&
&
&
&
2 &
849 &
100.0\% \tabularnewline
&
&
&
&
12 &
303 &
100.0\% \tabularnewline
\hline
\end{tabular}

\caption{Discovery of ordered firing sequences: Results (averaged 
over 100 datasets) of discovery of serial 
episodes with inter-event time constraints when multiple patterns are embedded. The 
columns in the table are essentially same as those in Table~\ref{tab:parallel}.}
\label{tab:serial}
\end{table}

\begin{figure}
\subfigure[]{\includegraphics[scale=0.75]{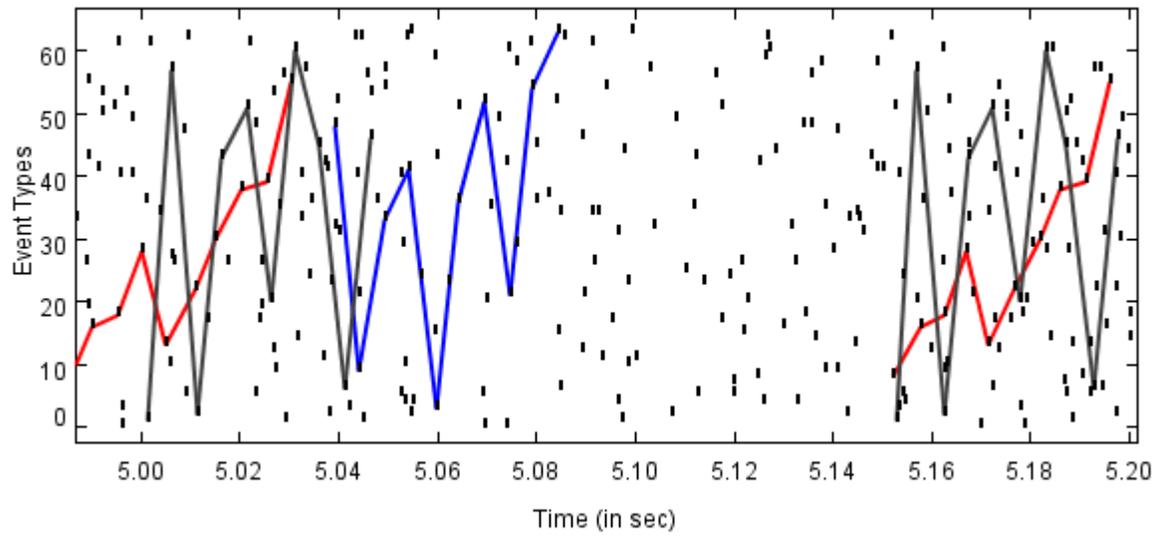}}

\subfigure[]{\includegraphics[scale=0.75]{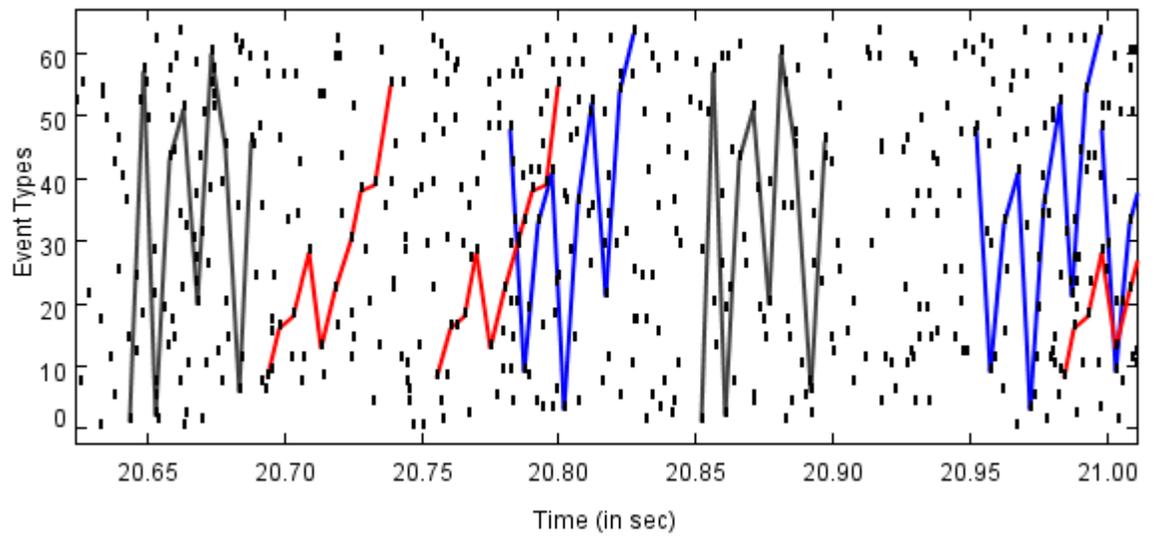}}

\caption{Some occurrences of ordered firing sequences in a typical dataset}
\label{fig:order-raster}
\end{figure}

In fig.~\ref{fig:order-raster} we illustrate some of the occurrences of the 
serial episodes in two different time windows on the data. 

\begin{table}
\centering
\begin{tabular}{|r|r|r|r|rrr|}
\hline
\multicolumn{ 1}{|c|}{Pattern} & \multicolumn{ 1}{|c|}{No. of} & 
\multicolumn{ 1}{|c|}{Episode } &  
\multicolumn{ 1}{|c|}{Time} & \multicolumn{ 3}{|c|}{Fraction of frequent episodes} \\

\multicolumn{ 1}{|c|}{Type} & \multicolumn{ 1}{|c|}{patterns} & 
\multicolumn{ 1}{|c|}{Type} &  
\multicolumn{ 1}{|c|}{(in sec)} & 
\multicolumn{ 3}{|c|}{that are part of embedded patterns} \\ \hline

\multicolumn{ 1}{|c|}{} & \multicolumn{ 1}{|c|}{} & 
\multicolumn{ 1}{|c|}{} &  
\multicolumn{ 1}{|c|}{} & Nodes & Min Count & \% Fraction \\
\hline
Syn-5 & 1 & Parallel &  1.042 & 1 & 965 & 31.2\% \\
1-(5)-1-(5)&   &   &      & 2 & 799 & 100.0\% \\
-1-(5)-1-(5)&   &   &      & 5 & 664 & 100.0\% \\
  &   & Serial (mod) &  0.861 & 1 & 663 & 16.7\% \\
  &   &   &     & 2 & 594 & 100.0\% \\
  &   &   &      & 8 & 192 & 100.0\% \\
  & 2 & Parallel &  1.232 & 1 & 955 & 62.5\% \\
  &   &   &   &    2 & 802 & 100.0\% \\
  &   &   &      & 5 & 659 & 100.0\% \\
  &   & Serial (mod) &  0.391 & 1 & 658 & 50.0\% \\
  &   &   &      & 2 & 614 & 100.0\% \\
  &   &   &      & 8 & 188 & 100.0\% \\
  & 3 & Parallel &  1.252 & 1 & 942 & 62.5\% \\
  &   &   &      & 2 & 794 & 100.0\% \\
  &   &   &      & 5 & 667 & 100.0\% \\
  &   & Serial (mod) &  0.39 & 1 & 666 & 50.0\% \\
  &   &      &   & 2 & 601 & 100.0\% \\
  &   &      &   & 8 & 181 & 100.0\% \\
\hline
Syn-5 & 1 & Parallel &  1.162 & 1 & 925 & 46.9\% \\
1-(5)-1-(5)&   &      &   & 2 & 780 & 100.0\% \\
-1-(5)-1-(5)&   &      &   & 5 & 635 & 100.0\% \\
-1-(5)-1-(5)&   & Serial (mod) &  0.631 & 1 & 634 & 30.0\% \\
  &   &   &      & 2 & 580 & 100.0\% \\
  &   &   &      & 12 & 74 & 100.0\% \\
\hline
\end{tabular}  
\caption{ Results of discovery for multiple synfire chain patterns. The general structure 
of the table is similar to the earlier ones.  
Here all parallel episodes are of size 5. We string together 4 or 6  such parallel 
episodes to make the synfire chain pattern and this is shown as pattern type in the table.  
We have used Parallel Episode expiry = 0.001 sec, and 
Serial Episode Inter-event time constraint = 0.004 to 0.006 sec. The table shows the two 
phases -- parallel episodes and modified serial episodes, separately both for time taken 
and for the percentage of discovered episodes that are part of embedded pattern.}
\label{tab:synfire-5}
\end{table}

\begin{table}
\centering
\begin{tabular}{|r|r|r|r|rrr|}
\hline
\multicolumn{ 1}{|c|}{Pattern} & \multicolumn{ 1}{|c|}{No. of} & \multicolumn{ 1}{|c|}{Episode} & 
 \multicolumn{ 1}{|c|}{Time} & 
\multicolumn{ 3}{|c|}{Fraction of frequent episodes} \\

\multicolumn{ 1}{|c|}{Type} & \multicolumn{ 1}{|c|}{patterns} & \multicolumn{ 1}{|c|}{Type} & 
 \multicolumn{ 1}{|c|}{(in sec)} & 
\multicolumn{ 3}{|c|}{that are part of embedded patterns} \\ \hline

\multicolumn{ 1}{|c|}{} & \multicolumn{ 1}{|c|}{} & \multicolumn{ 1}{|c|}{} & 
 \multicolumn{ 1}{|c|}{} & Nodes & Min Count & \% Fraction \\
\hline
Syn-4 & 1 & Parallel &  0.982 & 1 & 990 & 25.0\% \\
1-(4)-1-(4) &   &      &   & 2 & 848 & 100.0\% \\
-1-(4)-1-(4)  &   &      &   & 4 & 737 & 100.0\% \\
  &   & Serial (mod)  & 0.981 & 1 & 736 & 15.4\% \\
  &   &   &      & 2 & 664 & 100.0\% \\
  &   &   &      & 8 & 247 & 100.0\% \\
  & 2 & Parallel  & 1.032 & 1 & 936 & 50.0\% \\
  &   &   &      & 2 & 783 & 100.0\% \\
  &   &   &      & 4 & 697 & 100.0\% \\
  &   & Serial (mod)  & 0.561 & 1 & 696 & 40.0\% \\
  &   &   &      & 2 & 629 & 100.0\% \\
  &   &   &      & 8 & 226 & 100.0\% \\
  & 3 & Parallel  & 1.092 & 1 & 925 & 75.0\% \\
  &   &   &      & 2 & 781 & 100.0\% \\
  &   &   &      & 4 & 686 & 100.0\% \\
  &   & Serial (mod)  & 0.311 & 1 & 685 & 85.7\% \\
  &   &   &      & 2 & 617 & 100.0\% \\
  &   &   &      & 8 & 232 & 100.0\% \\
\hline
Syn-4 & 1 & Parallel  & 1.022 & 1 & 949 & 37.5\% \\
1-(4)-1-(4) &   &      &   & 2 & 798 & 100.0\% \\
-1-(4)-1-(4)  &     &   &   & 4 & 709 & 100.0\% \\
-1-(4)-1-(4)  &   & Serial (mod)  & 0.822 & 1 & 708 & 26.1\% \\
  &   &   &      & 2 & 630 & 100.0\% \\
  &   &   &      & 12 & 125 & 100.0\% \\
  & 2 & Parallel  & 1.052 & 1 & 934 & 75.0\% \\
  &   &   &      & 2 & 777 & 100.0\% \\
  &   &   &      & 4 & 682 & 100.0\% \\
  &   & Serial (mod)  & 0.381 & 1 & 681 & 85.7\% \\
  &   &   &      & 2 & 624 & 100.0\% \\
  &   &   &      & 12 & 111 & 100.0\% \\
\hline
Syn-4 & 1 & Parallel  & 1.031 & 1 & 936 & 50.0\% \\
1-(4)-1-(4)&   &      &   & 2 & 784 & 100.0\% \\
-1-(4)-1-(4)&   &      &   & 4 & 688 & 100.0\% \\
-1-(4)-1-(4)&   & Serial (mod)  & 0.711 & 1 & 687 & 40.0\% \\
-1-(4)-1-(4)&   &      &   & 2 & 620 & 100.0\% \\
  &   &   &      & 16 & 72 & 100.0\% \\
\hline
\end{tabular}  
\caption{Discovery of Synfire chain patterns where the parallel episodes are all of size 4.
We sring together 4, 6 or 8 such parallel episodes to make synfire chain patterns.  
Structure of table same as the previous one}
\label{tab:synfire-4}
\end{table}

\begin{figure}
\subfigure[]{\includegraphics[scale=0.75]{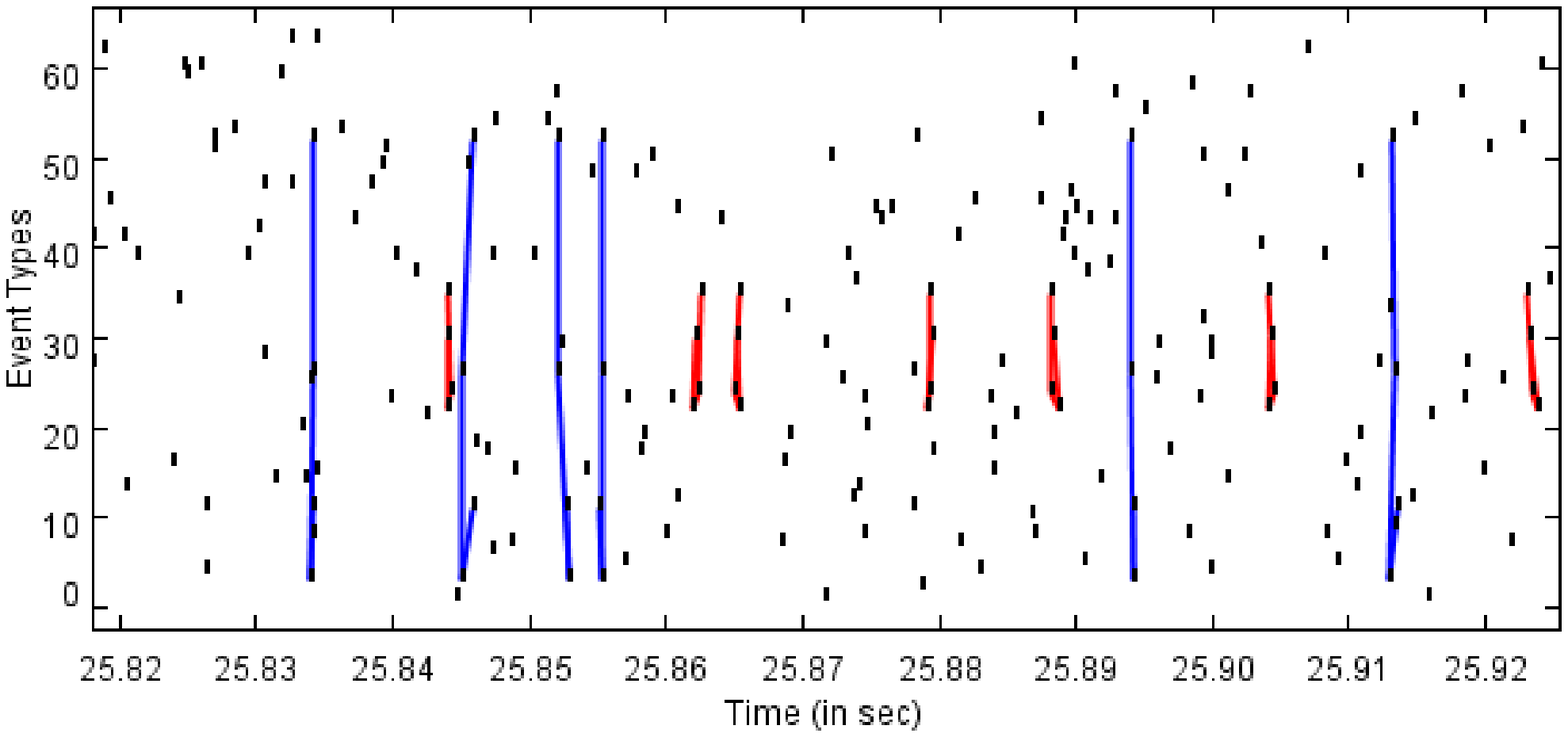}}

\subfigure[]{\includegraphics[scale=0.75]{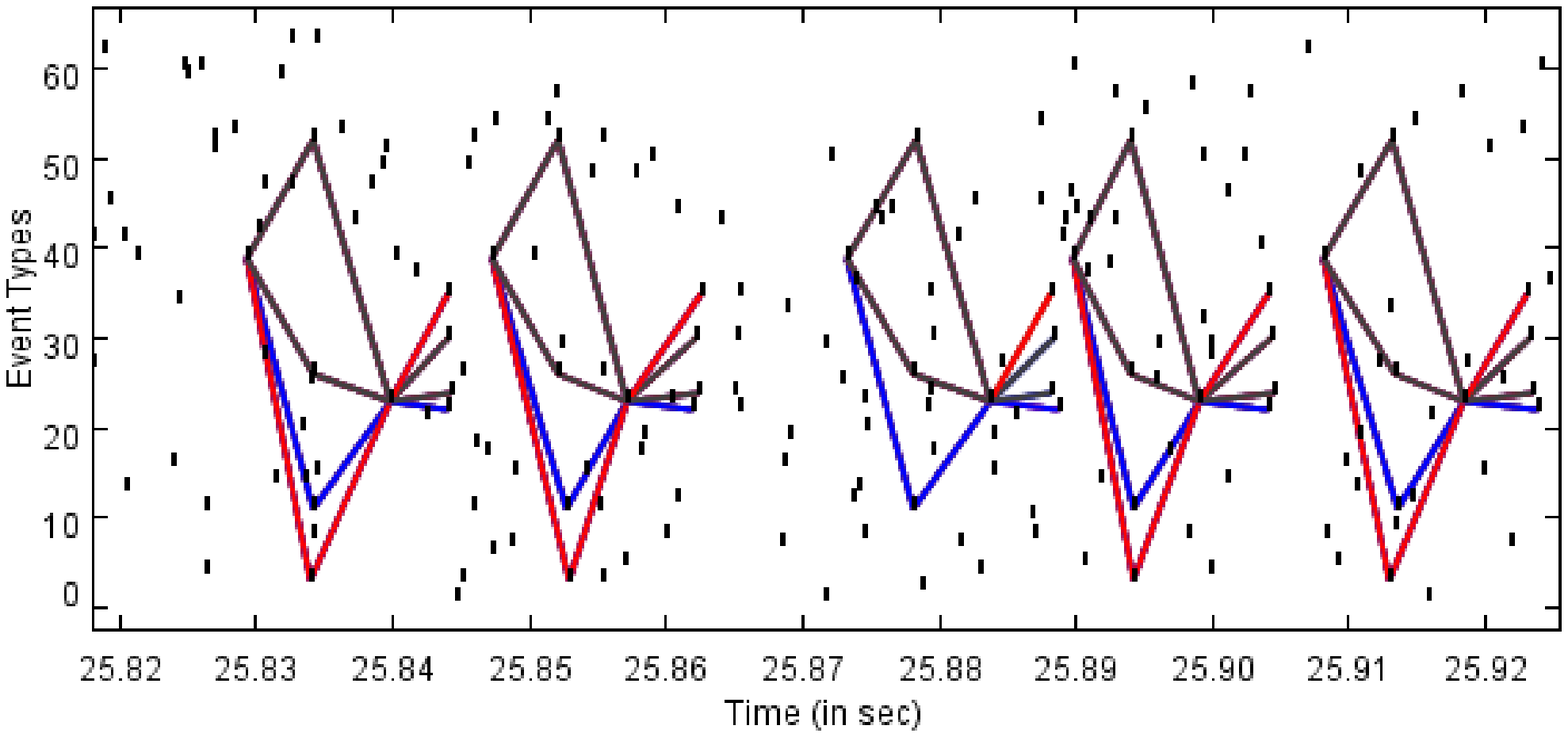}}

\caption{Some occurrances of  discovered Synfire chains}
\label{fig:synfire-raster}
\end{figure}

We have conducted similar experiments with multiple synfire chain patterns. 
Our method is equally effective in discovering Synfire chains.  
Table~\ref{tab:synfire-5} shows results obtained when each parallel episode 
inside the synfire chain has size 4. We have considered 4 and 6  such parallel 
episodes strung together to make the synfire chain and we have mebedded upto 2 such patterns. 
The table has the same structure as earlier tables and 
shows the type of synfire chain pattern, number of patterns embedded and the 
percentage of discovered patterns (of different sizes) 
which are part of the embedded patterns. Since a synfire chain pattern is composed of 
parallel episodes and serial episodes on modified data stream, we show the percentage 
of discovered patterns for these two cases separately. Table~\ref{tab:synfire-4} shows 
similar results for the case where each parallel episode in the synfire chain is 
of size 4. Here we string together 4, 6 or 8 such parallel episodes and we embed upto 3 
such patterns. From these tables, it is quite clear that our algorithms are very 
effective in unearthing synfire chain patterns even when multiple such patterns exist. 
Fig.~\ref{fig:synfire-raster} shows the discovered Synfire chains in  one set of data. 

\subsection{Assessing significance of frequent episodes discovered}

The empirical results presented earlier show that if we generate 
spike data using special embedded 
patterns in it then our algorithms can detect them. That is, if the spike 
data is generated by a system of interconnected neurons with a few strong 
excitatory connections, then the frequent episodes we discover 
clearly bring out the connection pattern in the network. 
However, these results do not answer the 
question: if the algorithm detects some frequent episodes what confidence do we have 
that they correspond to some patterns in the underlying neural system. 
This is a question regarding the statistical significance of the discovered 
frequent episodes. That is, we need to ask how high should the frequency of an 
episode  be so that, with a high confidence, we can assert that the 
connectivity pattern implied by the episode is, in some sense, characteristic 
of the system generating the data. 

To answer this question we have to essentially choose a {\em null hypothesis} that asserts 
that there is no `structure' in the system generating the data. Then we need to 
calculate the probability that an episode of a given size would have a given frequency 
in the data generated by such a model and this will tell us what is the chance of a 
discovered frequent episode coming up  
by chance in `random' data and hence tells us the statistical 
significance of the discovered frequent episodes. This can also allow us to calculate 
the frequency threshold so that all frequent episodes (with frequency above this 
threshold) are statistically significant at a given level of confidence. Such an analysis 
is presented for the case of serial episodes without any temporal constraints under the 
null hypothesis that the event stream is generated by an {\em iid} process 
in \cite{Srivats2005}. 

That analysis is not directly applicable for the multineuronal spiking data application 
because we have temporal constraints on episode occurrances here. But, more importantly, 
a null hypothesis of an {\em iid} process generating the spike data is not really attractive 
here. For example, even if we can reject the null hypothesis that spike trains are 
produced by independent Poisson processes, it does not mean that the system generating the 
data has strong correlations or connectivity patterns as indicated by our episodes. 

We like to note here that all current methods of spike data analysis, whenever they 
consider issues of statistical significance, deal with a null hypothesis of {\em iid} 
processes generating spike data. One notable exception is the work in \cite{amarasing-thesis} 
where more complicated null hypotheses are considered. However, this work does not 
deal with finding useful patterns in spike data; the objective of the 
analysis there is to determine the time scale at which exact times of spikes may carry 
useful information as opposed to all information being carried by only the spiking 
frequency.  

Here we want to consider a composite null hypothesis in which we 
include  not only {\em iid} processes, but also other models 
for interdependent neurons without any specific strong connectivity patterns or strong 
predispositions for coordinated firing. It is 
difficult to capture all such models in an analytically tractable mathematical 
formulation. Hence,  we take the approach of capturing our null hypothesis in 
a simulation model and estimate the relevant probabilities by generating many random 
data sets from such a model. (This approach is similar in spirit to the so called 
`jitter' method \cite{date2001}). 

We generate our random data sets using essentially three different types of models. 
For the first one, we use the same simulator as described 
earlier; but we allow only the random interconnections (with weights of 
interconnections uniformly distributed over $[-c,\  c]$). This will capture models where 
the poisson processes representing spikes by different neurons are interdependent 
(with the firing rate of a neuron being dependent on spikes output by other neurons)  
but without any bias for some specific interconnectivity pattern or coordinated 
firing. Next, we  
generate data sets by assuming that different neurons generate spikes as independent 
Poisson processes by simply choosing random fixed rates for the neurons. In this, we 
also include cases where many neurons can have the same firing rate. For this, we fix 
five or ten different random firing rates and randomly assign each neuron to have 
one of these firing rates. For our third type of data sets, we include 
models where rates of firing by neurons change; but without any relation to firing by 
other neurons. For this we choose random firing rates for neurons and  at  
$50 \Delta{t}$ time steps we randomly perturb the firing rate. Here also we include the 
case where firing rates of some random subsets of neurons are all tied together.  

Thus our null hypothesis includes 
models where different neurons could be {\em iid} Poisson processes, or 
inhomogeneous Poisson processes where the firing rates may be correlated but the rate 
is not dependent on firing of other neurons. In addition, 
our null hypothesis also includes models where rates of firing change 
based on spikes output by other neurons but without any bias for specific strong 
interconnectivity patterns. We feel that this is a large enough set of models to 
consider in the null hypothesis. If, based on our episode frequencies, we can 
reject the null hypothesis, then, it clearly demonstrates that episodes with 
sufficiently high frequency can not come about unless there is a bias or interdependence 
in the underlying neural system for coordinated firing by some groups of neurons. 

By generating many data sets under the models in our null hypothesis 
and calculating frequencies of 
episodes of different sizes we now show that 
 it is highly unlikely to have long frequent episodes if the data generation model does 
not have any specifiic biases. 

The specific random data sets are as follows. All data sets are from 
a 26 neuron system. 
 We generated ten sets of random interconnection weights. In one half of these, the weights 
are chosen with a uniform distribution over $[-0.5, \ 0.5]$ and  in the 
other half the weights are uniformly distributed over $[-0.5 \  0.5]$. 
With each set of random weights,  
 we generated ten sets of data (25 000 spikes) 
by running the simulator with those weights. The normal firing rate (i.e., the rate of 
firing when input is zero) is set at 20 Hz.  Thus 
we have 100 data sets in which, while neurons fire with input dependent firing 
rates, there are no special causative connections. We generated another 25 data sets 
(of 25,000 spikes each) where each neuron had a fixed firing rate chosen from a 
uniform distribution over $[10 Hz, \ 30 Hz]$. In another 25 data sets, we have  
five different firing rates chosen randomly 
from the same interval and each neuron is randomly 
assigned to one of these firing rates. In another 25 data sets we randomly choose a 
new firing rate for each neuron from the same inteval every $50 \Delta{t}$ time units. 
Here different neurons are independent inhomogeneous Poisson processes. Finally, in 
another 25 data sets, we randomly divide the neurons into five groups. All neurons 
in a group would have the same firing rate. The firing rates are updated every 
$50 \Delta{t}$ time units by choosing from a uniform distribution over the same interval 
as earlier. Thus, we have a total of 200 data sets. In half of them, the firing rate 
of any neuron is a function of the actual outputs of other neurons connected to it 
though all interconnections weights are random with zero mean. Thus, though the 
neurons are interdependent, there are no biases for any specific connection pattern. 
In the other half of the data sets the firing rates of neurons are random, fixed or 
randomly changed, some of the neurons may be correlated in the sense of having the same 
fixed or varying firing rate, but firing rates are not dependent on outputs of other 
neurons. We also note here that in all cases we have 
chosen the random firing rates in such a way that the average firing rate of the total 
system is roughly same as that in the experiments described in the earlier subsections. 

In each of these data sets, 
We  discover serial and parallel episodes (with the usual temporal 
constraints)  of size upto 10 with a 
frequency threshold of zero so that we get frequencies for all episodes. 
We then compare the maximum frequencies  (averaged over all 
data sets) of episodes of 
different sizes obtained from this random 
data sets to the minimum frequencies observed for same size episodes on data sets 
with patterns embedded in them. For this comparison we have generated another 
twenty data sets (using the earlier simulator) where a large episode (serial or 
parallel as needed) is embedded and the average firing rate is same as that in the 
random data sets. For the random data case, we show the results as two parts. First 
part corresponds to the 100 data sets where neurons have spike-input dependent 
firing rates. The second part corresponds to the 100 data sets where fixed or varying 
random firing rates are chosen for the neurons.

\begin{table}
\centering
\begin{tabular}{|r|r|r|r|} \hline
Size & \multicolumn{ 3}{|c|}{Max. Episode Frequency} \\ \hline
  & Avg & Max & Min \\ \hline
1-Node & 1084.81 & 1165.00 & 1038.00 \\ \hline
2-Node & 60.21 & 76.00 & 54.00 \\ \hline
3-Node & 6.52 & 9.00 & 5.00 \\ \hline
4-Node & 2.03 & 3.00 & 2.00 \\ \hline
5-Node & 0.06 & 2.00 & 0.00 \\ \hline
6-Node & 0.00 & 0.00 & 0.00 \\ \hline
7-Node & 0.00 & 0.00 & 0.00 \\ \hline
8-Node & 0.00 & 0.00 & 0.00 \\ \hline
9-Node & 0.00 & 0.00 & 0.00 \\ \hline
10-Node & 0.00 & 0.00 & 0.00 \\ \hline
  & \multicolumn{ 3}{|c|}{Sample size = 100} \\ \hline
\end{tabular}
\caption{Statistics for parallel episode mining on random spike sequences generated using 
the model of interdependent neurons with random interconnections.  
(Parallel Episode Expiry constraint = 0.001 sec). For different sizes of episodes, we show 
the maximum frequency of any episode of that size. Each entry shows the maximum, minimum 
and average values of the maximum frequency. These statistics are obtained from a 
sample size of 100 data sets}
\label{tab:sig-par-ns-conn}
\end{table}

\begin{table}
\centering
\begin{tabular}{|r|r|r|r|} \hline
Size & \multicolumn{ 3}{|c|}{Max. Episode Frequency} \\ \hline
  & Avg & Max & Min \\ \hline
1-Node & 1249.00 & 1513.00 & 1004.00 \\ \hline
2-Node & 76.55 & 111.00 & 52.00 \\ \hline
3-Node & 7.62 & 12.00 & 5.00 \\ \hline
4-Node & 2.12 & 3.00 & 0.00 \\ \hline
5-Node & 0.04 & 2.00 & 0.00 \\ \hline
6-Node & 0.00 & 0.00 & 0.00 \\ \hline
7-Node & 0.00 & 0.00 & 0.00 \\ \hline
8-Node & 0.00 & 0.00 & 0.00 \\ \hline
9-Node & 0.00 & 0.00 & 0.00 \\ \hline
10-Node & 0.00 & 0.00 & 0.00 \\ \hline
  & \multicolumn{ 3}{|c|}{Sample size = 100} \\ \hline
\end{tabular}  
\caption{Statistics for parallel episode mining on random spike sequences generated using 
random (fized or varying) firing rates for neurons.  
(Parallel Episode Expiry constraint = 0.001 sec). For different sizes of episodes, we show 
the maximum frequency of any episode of that size. Each entry shows the maximum, minimum 
and average values of the maximum frequency. These statistics are obtained from a 
sample size of 100 data sets}
\label{tab:sig-par-ns-no-conn}
\end{table}

\begin{table}
\centering
\begin{tabular}{|r|r|r|r|} \hline
Size & \multicolumn{ 3}{|c|}{Min. Episode Frequency} \\ \hline
  & Avg & Max & Min \\ \hline
1-Node & 978.80 & 1074.00 & 933.00 \\ \hline
2-Node & 822.70 & 913.00 & 777.00 \\ \hline
3-Node & 763.35 & 853.00 & 714.00 \\ \hline
4-Node & 711.20 & 801.00 & 664.00 \\ \hline
5-Node & 657.56 & 702.00 & 620.00 \\ \hline
6-Node & 616.25 & 671.00 & 585.00 \\ \hline
7-Node & 571.50 & 598.00 & 543.00 \\ \hline
8-Node & 537.08 & 570.00 & 504.00 \\ \hline
9-Node & 493.50 & 525.00 & 468.00 \\ \hline
10-Node & 464.88 & 494.00 & 433.00 \\ \hline
  & \multicolumn{ 3}{|c|}{Sample size = 20} \\ \hline
\end{tabular}  
\caption{Statistics for parallel episode mining on  spike sequence data with 
embedded patterns in it. 
(Parallel Episode Expiry constraint = 0.001 sec). For different sizes of episodes, we show 
the minimum frequency of any episode of that size which is a subepisode of the embedded 
pattern.  Each entry shows the maximum, minimum 
and average values of the minimum frequency. These statistics are obtained from a 
sample size of 20 data sets}
\label{tab:sig-par-pattern}
\end{table}

\begin{figure}[!htb]
\centering
\epsfig{file=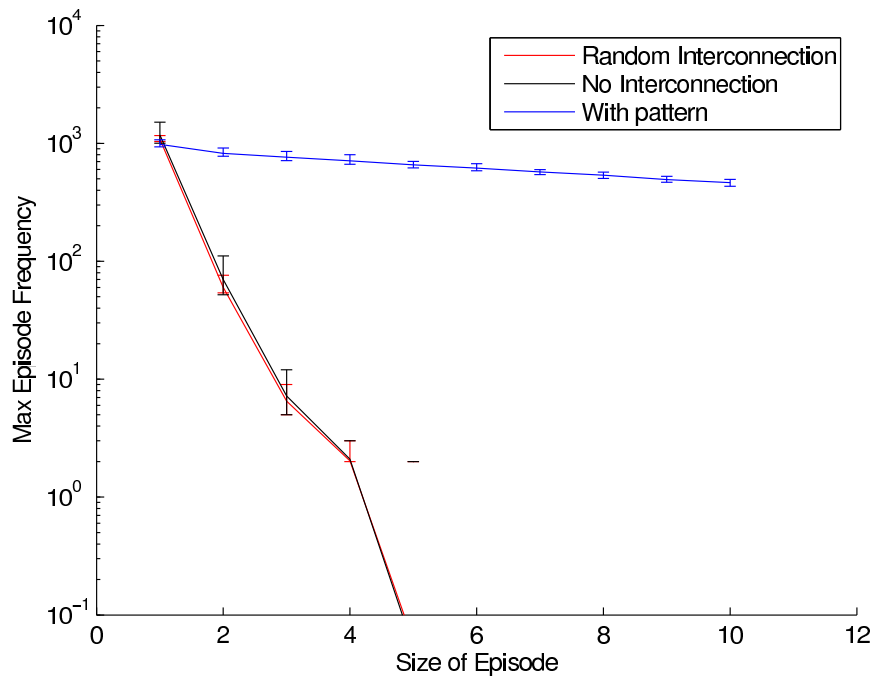,height=3.5in,width=4.5in}
\caption{Differences in episode frequencies in noise sequences and sequences with 
patterns. All are parallel episodes with expiry time constraint of 0.001 sec. For the 
case of data with patterns, we show minimum frequency for any episode that is part 
of the embedded pattern and the error bars show the range of variation. The figure 
clearly shows that it is extremely unlikely to have episodes of size 3 or more with 
appreciable frequency `by chance'. }
\label{fig:parallel-sig}
\end{figure}

Tables~\ref{tab:sig-par-ns-conn} -- \ref{tab:sig-par-pattern} show the results 
obtained in case of parallel episodes. Table~\ref{tab:sig-par-ns-conn} shows the 
maximum observed frequency of episodes for various sizes in case of the random data 
obtained from our model of interdependent neurons but with random interconnection weights. 
Table~\ref{tab:sig-par-ns-no-conn} shows the same for the case of data generated using 
different kinds of random firing rates for neurons as explained earlier. These tables 
show statistics obtained from a sample size of 100 data sets each. These numbers are to be 
compared with those in Table~\ref{tab:sig-par-pattern} minimum observed frequency of 
episodes that are part of embedded patterns for different sizes. These numbers are obtained 
from ours simulation model with one large parallel episode embedded. These are statistics 
obtained from a sample of 20 data sets. Fig.~\ref{fig:parallel-sig} shows the plot of 
maximum frequencies of episodes in noise sequences and minimum frequency of relevant 
episodes in sequences with patterns versus episode size. For the plot showing 
minimum frequencies for the correct episodes in data with patterns, we show the 
observed variation in minimum frequency as an error bar on the figure. 

From the table it can be seen that, even for size 2, the maximum frequency of an episode 
in the random data is very small. From size 3 onwards, all episodes have frequency less 
than 10 in the random data. On the other hand, when the data contains patterns, 
even the minimum observed frequencies of that size 
episodes (which are part of the embedded pattern or ground truth) 
 are about two orders of magnitude larger. We also note here that the frequencies of 
1-node episodes are comparable in the random and patterned data sets which is due to the 
fact that we have ensured that the average firing rates of neurons in both sets of data 
are same. These results provides sufficient statistical 
justification that it is higly unlikely to have long episodes with appreciable frequencies 
if the data source does not have the necessary bias. 

\begin{table}
\centering
\begin{tabular}{|r|r|r|r|} \hline
Size & \multicolumn{ 3}{|c|}{Max. Episode Frequency} \\ \hline
  & Avg & Max & Min \\ \hline
1-Node & 1084.81 & 1165.00 & 1038.00 \\ \hline
2-Node & 70.51 & 84.00 & 59.00 \\ \hline
3-Node & 9.22 & 13.00 & 8.00 \\ \hline
4-Node & 3.30 & 4.00 & 3.00 \\ \hline
5-Node & 2.02 & 3.00 & 2.00 \\ \hline
6-Node & 1.24 & 2.00 & 0.00 \\ \hline
7-Node & 0.04 & 2.00 & 0.00 \\ \hline
8-Node & 0.02 & 2.00 & 0.00 \\ \hline
9-Node & 0.00 & 0.00 & 0.00 \\ \hline
10-Node & 0.00 & 0.00 & 0.00 \\ \hline
  & \multicolumn{ 3}{|c|}{Sample size = 100} \\ \hline
\end{tabular}  
\caption{Statistics for serial episode mining on random spike sequences generated using 
the model of interdependent neurons with random interconnections.  
(Serial Episode Inter-event time constraint is 0.004-0.006 sec)
For different sizes of episodes, we show 
the maximum frequency of any episode of that size. Each entry shows the maximum, minimum 
and average values of the maximum frequency. These statistics are obtained from a 
sample size of 100 data sets}
\label{tab:sig-ser-ns-conn}
\end{table}

\begin{table}
\centering
\begin{tabular}{|r|r|r|r|} \hline
Size & \multicolumn{ 3}{|c|}{Max. Episode Frequency} \\ \hline
  & Avg & Max & Min \\ \hline
1-Node & 1249.00 & 1513.00 & 1004.00 \\ \hline
2-Node & 78.22 & 113.00 & 51.00 \\ \hline
3-Node & 10.22 & 15.00 & 7.00 \\ \hline
4-Node & 3.54 & 6.00 & 3.00 \\ \hline
5-Node & 2.10 & 3.00 & 2.00 \\ \hline
6-Node & 1.24 & 2.00 & 0.00 \\ \hline
7-Node & 0.10 & 2.00 & 0.00 \\ \hline
8-Node & 0.00 & 0.00 & 0.00 \\ \hline
9-Node & 0.00 & 0.00 & 0.00 \\ \hline
10-Node & 0.00 & 0.00 & 0.00 \\ \hline
  & \multicolumn{ 3}{|c|}{Sample size = 100} \\ \hline
\end{tabular}  
\caption{Statistics for serial episode mining on random spike sequences generated using 
random (fized or varying) firing rates for neurons.  
(Serial Episode Inter-event time constraint = 0.004-0.006 sec)
For different sizes of episodes, we show 
the maximum frequency of any episode of that size. Each entry shows the maximum, minimum 
and average values of the maximum frequency. These statistics are obtained from a 
sample size of 100 data sets}
\label{tab:sig-ser-ns-no-conn}
\end{table}

\begin{table}
\centering
\begin{tabular}{|r|r|r|r|} \hline
Size & \multicolumn{ 3}{|c|}{Min. Episode Frequency} \\ \hline
  & Avg & Max & Min \\ \hline
1-Node & 967.80 & 1032.00 & 916.00 \\ \hline
2-Node & 845.65 & 903.00 & 798.00 \\ \hline
3-Node & 734.55 & 770.00 & 702.00 \\ \hline
4-Node & 647.30 & 679.00 & 608.00 \\ \hline
5-Node & 576.06 & 615.00 & 548.00 \\ \hline
6-Node & 515.88 & 545.00 & 482.00 \\ \hline
7-Node & 466.33 & 487.00 & 448.00 \\ \hline
8-Node & 423.58 & 447.00 & 405.00 \\ \hline
9-Node & 385.25 & 395.00 & 376.00 \\ \hline
10-Node & 353.88 & 368.00 & 345.00 \\ \hline
  & \multicolumn{ 3}{|c|}{Sample size = 20} \\ \hline
\end{tabular}  
\caption{Statistics for parallel episode mining on  spike sequence data with 
embedded patterns in it. 
(Serial Episode Inter-event time constraint = 0.004-0.006 sec)
For different sizes of episodes, we show 
the minimum frequency of any episode of that size which is a subepisode of the embedded 
pattern.  Each entry shows the maximum, minimum 
and average values of the minimum frequency. These statistics are obtained from a 
sample size of 20 data sets}
\label{tab:sig-ser-pattern}
\end{table}

\begin{table}
\centering
\begin{tabular}{|r|r|r|r|} \hline
Size & \multicolumn{ 3}{|c|}{Max. Episode Frequency} \\ \hline
  & Avg & Max & Min \\ \hline
1-Node & 1249.00 & 1513.00 & 1004.00 \\ \hline
2-Node & 81.48 & 115.00 & 56.00 \\ \hline
3-Node & 11.62 & 20.00 & 8.00 \\ \hline
4-Node & 4.41 & 6.00 & 3.00 \\ \hline
5-Node & 2.87 & 4.00 & 2.00 \\ \hline
6-Node & 2.05 & 3.00 & 2.00 \\ \hline
7-Node & 2.00 & 2.00 & 2.00 \\ \hline
8-Node & 1.46 & 2.00 & 0.00 \\ \hline
9-Node & 0.44 & 2.00 & 0.00 \\ \hline
10-Node & 0.10 & 2.00 & 0.00 \\ \hline
  & \multicolumn{ 3}{|c|}{Sample size = 100} \\ \hline
\end{tabular}
\caption{Statistics for serial episode mining with inter-event time constraint  discovery 
on data sets generated with (fixed or varying) random firing rates. The set of possible 
inter-event time constraints are: 0.002-0.004 sec,0.004-0.006 sec,0.006-0.008 sec. As before 
we show statistics of maximum observed frequencies for episodes of different sizes}
\label{tab:sig-ser-ns-int}
\end{table}

\begin{figure}[!htb]
\centering
\epsfig{file=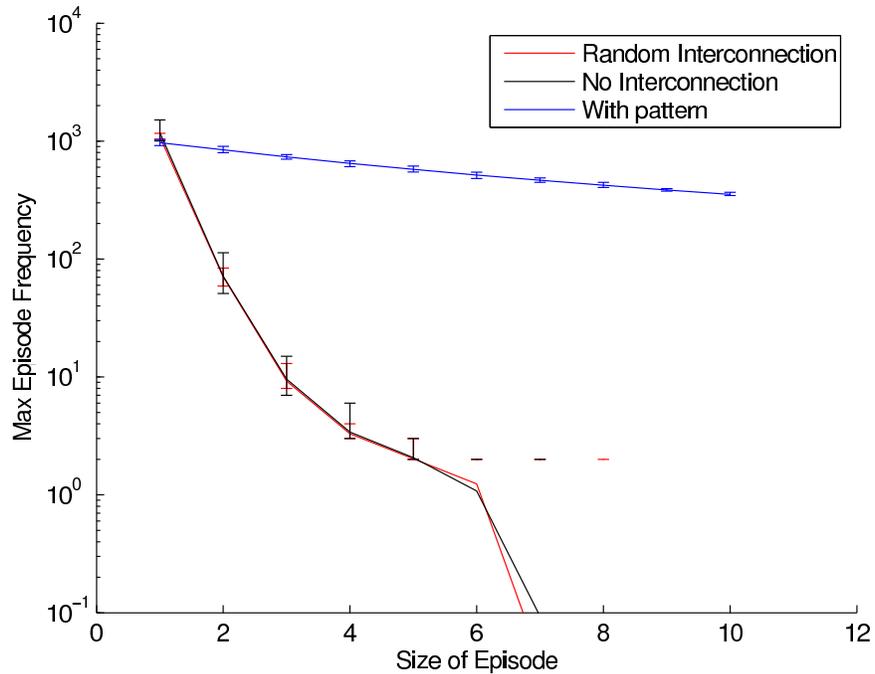,height=3.5in,width=4.5in}
\caption{Differences in episode frequencies in noise sequences and sequences with 
patterns. All are serial episodes with inter-event time constraint of 0.004--0.006 sec. 
For the 
case of data with patterns, we show minimum frequency for any episode that is part 
of the embedded pattern and the error bars show the range of variation. The figure 
clearly shows that it is extremely unlikely to have episodes of size 3 or more with 
appreciable frequency `by chance'. }
\label{fig:serial-sig}
\end{figure}

Tables~\ref{tab:sig-ser-ns-conn} and \ref{tab:sig-ser-ns-no-conn} show the results 
obtained for serial episode mining for the two different kinds of random spike 
sequences as earlier and Table~\ref{tab:sig-ser-pattern} shows the results for serial 
episode mining on data that contains patterns in it. Once again, it is seen that 
in random data the maximum frequency of an episode falls rapidly with size of episode and 
it is less than 10 for size 3 onwards. On the other hand, if we generate data with 
specific pattern of interconnections then even the minimum frequency of episodes that 
are part of the mebedded pattern is about two orders of magnitude higher. In all these 
tables we have used an inter-event time constraint of 0.004--0.006 sec. For random data 
generated through interconnected neurons model, this is reasonable because we have a 
synaptic delay of 0.005 sec. However, one may argue that in data generated through 
random firing rates, there may be episodes with higher frequencies if we consider 
other inter-event time constraints. So, on these sets of random data, we have used 
our mining algorithm that can automatically detect the best possible inter-event 
constraint from a given set of constraints. These results are shown in 
Table~\ref{tab:sig-ser-ns-int}. It is easily seen that even in this case, the frequencies 
of episodes fall of very rapidly with episode size. Fig.~\ref{fig:serial-sig} shows the 
plot of episode frequencies versus size in case of serial episodes in noise data as well as 
in data with patterns embedded in it. 

All the above results clearly demonstrate that it is extremely unlikely to have large size 
episodes with any appreciable frequencies in random spike data. 
For example, from the above results we can 
conclude that the probability of having an episode of size greater than 2 with a 
frequency of 300 under the null hypothesis 
is less than 0.005 because not even once in 200 samples from the 
Null hypothesis did we get an episode of this frequency. This then is the p-value for 
asserting that an episode of frequency above 300 is significant. 
Of course, given the vast difference between frequencies of episodes in random and patterned 
data, such p-values are really not important. 
From the tables, it is also clear that frequency threshold of 250 or 300 (in data of this 
length) brings out only significant patterns.

Given a specific data set (which may be obtained through experiments on neural cultures) one 
can use the above method for assessing significance of discovered episodes as follows. 
From the data we estimate average firing rates of individual neurons and also firing rates 
averaged over windows of appropriate width. We use these to set the random firing rates 
as well as the variations in firing rates in our models for generating the random data 
sets. Then we generate many random data sets of the same length as the given data and discover
episodes with the same temporal constraints as in the real data. Then, by comparing 
frequencies as above, we can say which of the discovered episodes are significant. As a 
matter of fact, using the distribution of frequencies in the random data, we can set the 
frequency thresholds for our algorithms to discover significant episodes in the real data.

\subsection{\label{sec:Analysis-of-multi-neuron}Analysis of multi-neuron data obtained 
through Calcium Imaging}

\begin{figure}
\centering
\includegraphics[scale=0.5]{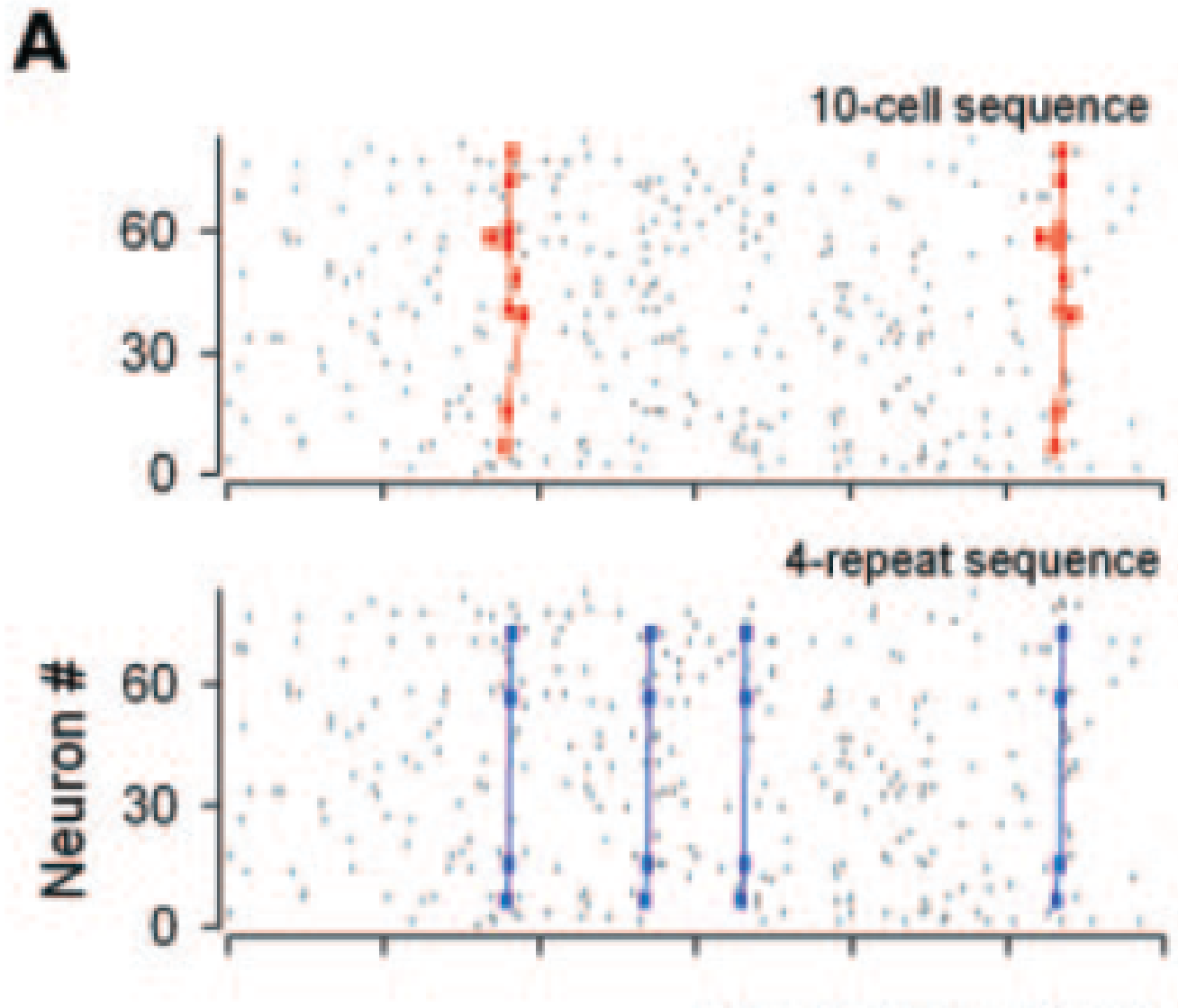}
\caption{\label{fig:Rafa-paper}Repeated motifs during sequential reactivation
of identical cells (A) Set of neurons with precise sequences of calcium
transients (V1 slice). Ten cells reactivated with exact timings between
their transients (top panel). In the same raster plot, a four-cell
sequence is reactivated four times (middle panel). This four-cell
sequence also acted as a part of the 10-cell sequence. Bottom panel
shows all sequences detected in the same raster plot.}
\end{figure}

\begin{figure}[!htb]
\centering
\subfigure[Frequent parallel episodes of size 10 satisfying expiry constraint = 10 time units]{\epsfig{file=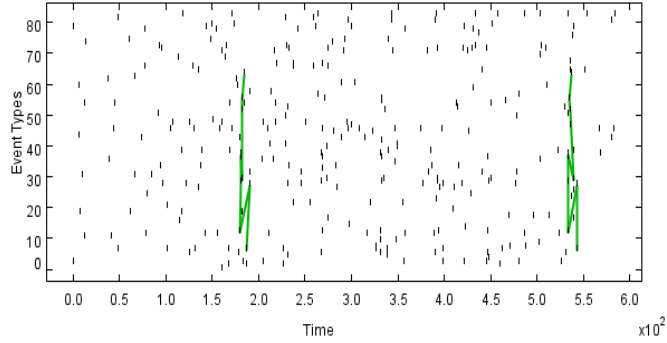,width=3.5in,height=2in}}
\subfigure[Frequent serial episodes of size 4 satisfying inter-event interval constraint = 10 time units]{\epsfig{file=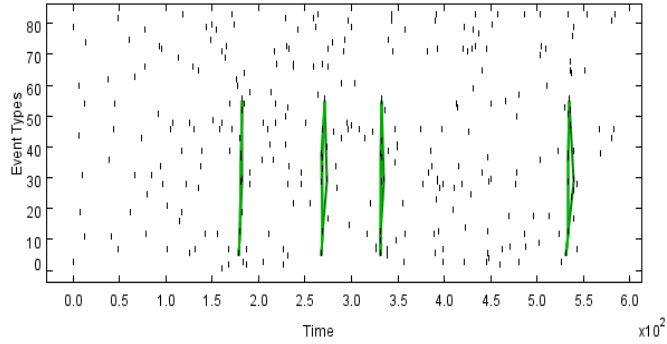,width=3.5in,height=2in}}
\caption{\label{fig:our-results}Frequent episodes discovered using our algorithms on real data.}
\end{figure}

In this section we describe results obtained on data sets collected
from experiments on neural ensembles. This  data set is from Dr.
Rafael Yuste's\footnote{We are extremely grateful to Dr. Rafael Yuste for sharing the calcium imaging data with us.} lab in Dept. of Biological Sciences, Columbia University,
New York and the results of these experiments are reported in \cite{Ikegaya2004}.
In this experiment, 
the data is obtained through calcium imaging
technique. 
In \cite{Ikegaya2004}, Ikegaya et. al. analyzed how neural activity
propagates through cortical networks. They found precise repetitions
of spontaneous patterns. These patterns repeated after minutes maintaining
millisecond accuracy. In Fig. 3A of \cite{Ikegaya2004}, such patterns 
are shown in raster plots by connecting the spikes that are part of
an occurrence.

Fig.~\ref{fig:Rafa-paper} shows a figure reproduced from \cite{Ikegaya2004} along 
with its original caption. 
In Fig.~\ref{fig:our-results}, we show results obtained on the same
calcium imaging data set using frequent episode discovery algorithms.
Fig.~\ref{fig:our-results}~(a) shows two occurrences of a 10-node
parallel episode discovered 
with expiry time constraint $T_{X}=10$ time units. Fig.~\ref{fig:our-results}~(b)
shows four occurrences of a 4-node serial episode discovered 
 with inter-event constraint
of 0 to 10 time units. It is seen that the results obtained
using frequent episode discovery match with those presented in \cite{Ikegaya2004} 
by comparing figures \ref{fig:Rafa-paper} with \ref{fig:our-results}.
Also, the time needed by our algorithm is much smaller because in 
\cite{Ikegaya2004}, they use a counting technique that can not control the 
combinatorial explosion. 
This result brings out the utility of our data mining 
 technique in terms of both effectiveness and efficiency.

\subsection{Analysis of multi-neuron data obtained through multi-electrode array experiments}
\label{sec:multi-neuron-mea-data}

In this subsection we present some of the results we obtained with our algorithms 
on multi-neuronal data obtained through multi-electrode array experiments.\footnote{We 
are grateful to Prof. Steve Potter, Georgia Institute of Technology and Emory 
University, Atlanta, USA, for providing the data and for many useful discussions on 
analyzing this data.} The data is obtained from dissociated cultures of cortical 
neurons grown on multi-electrode arrays. This is an extremely rich set of data where 
58 cultures of varying densities are followed for five weeks. Everyday, the spontaneous 
activity as well as 
 stimulated activity of each culture is recorded for different time durations. 
(See \cite{Potter2006} for the details of experiments, nature of data, trends observed etc.). 
Since data was recorded from each culture for many days, one can presumably infer development 
of connections also. Here we only present a few of the results we obtained from analyzing the 
spontaneous data from these 
cultures, to illustrate the utility of our temporal datamining techniques. 

In these dissociated cortical cultures, there is a lot of spontaneous activity including 
many cycles of network-wide bursts \cite{Potter2006}. Thus, patterns of coordinated 
firing by groups of neurons, even when they exist, would be rare in the sense that the 
spikes which form the coordinated activity constitute only 
 a small fraction of the total number of 
spikes output by the system. Thus, simple cross correlation based methods are not very 
effective in unearting coordinated firing patterns. Using our algorithm for serial 
episode discovery under inter-event constraints, we are able to obtain some frequent 
episodes which remain frequent for a large number of days with increasing trend in 
frequency.

\begin{figure}[!htb]
\centering
\epsfig{file=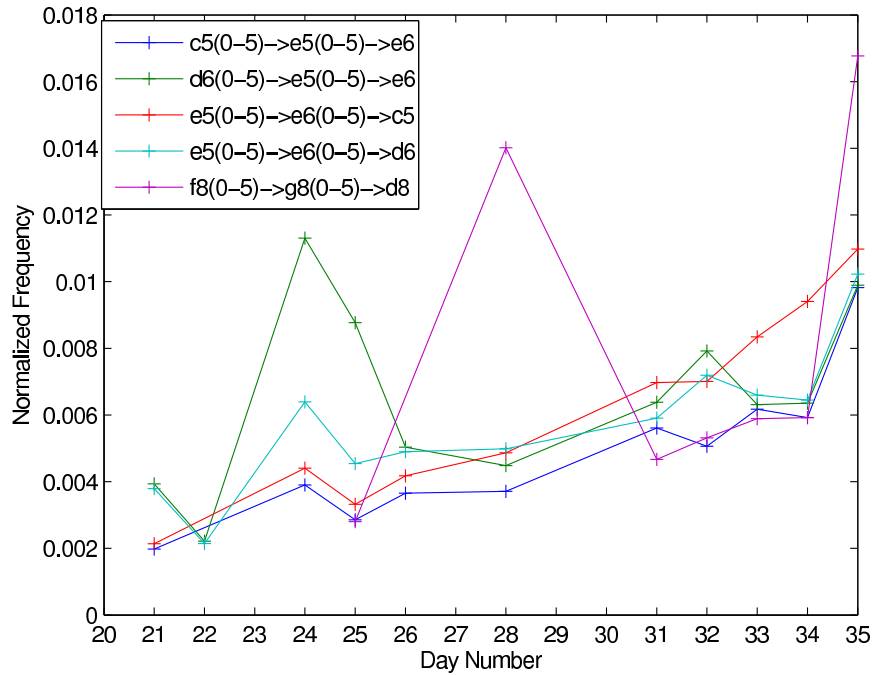,height=3.5in,width=4.5in}
\caption{Some frequent Serial Episodes discovered from multi-electrode array data 
from \cite{Potter2006}. We plot the normmalized frequency versus the age in days of the 
culture. These are from data of culture 2-1.} 
\label{fig:potter-norm-freq}
\end{figure}

\begin{figure}[!htb]
\centering
\epsfig{file=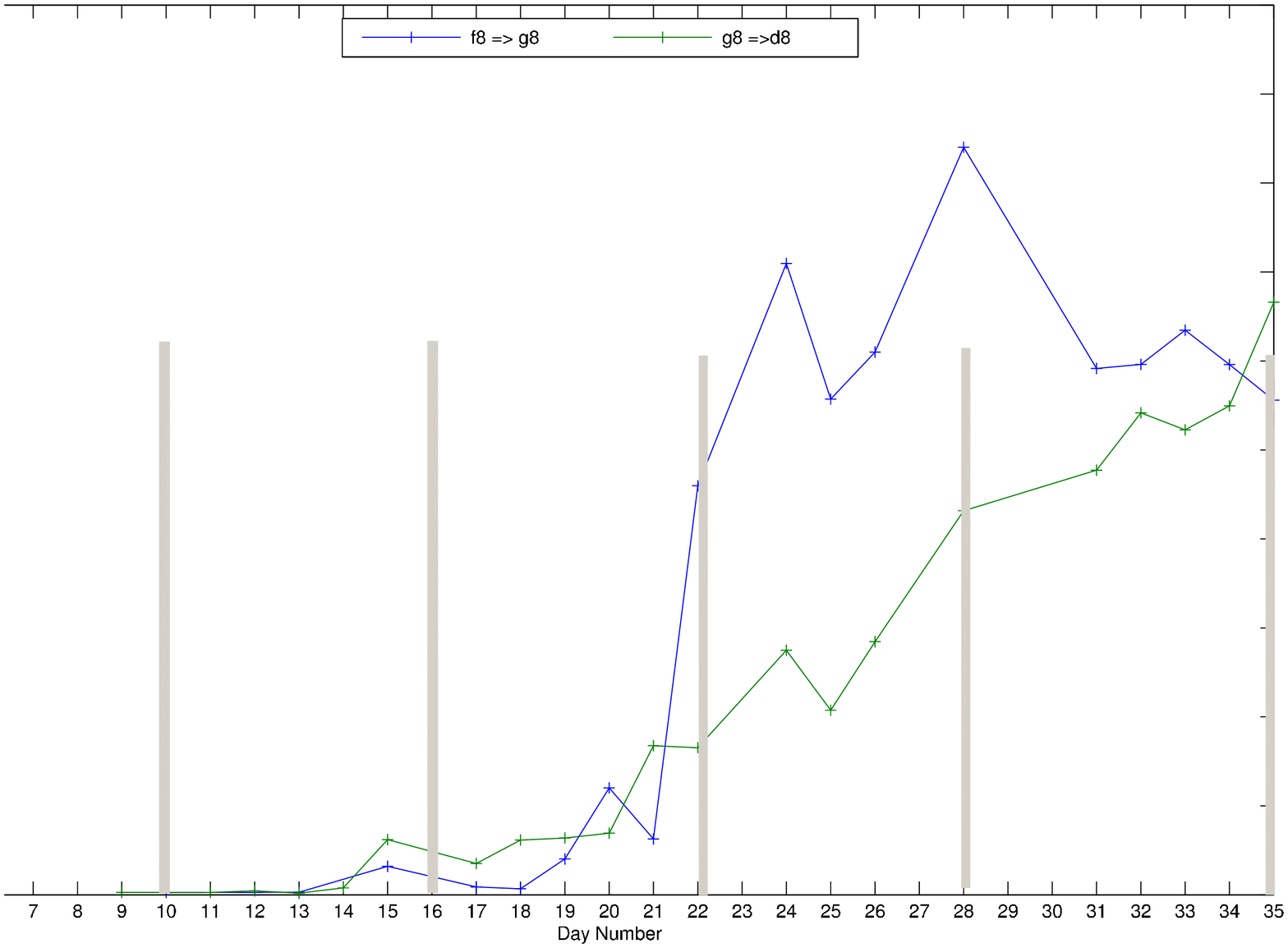,height=3.5in,width=4.5in}
\caption{Confidence scores for subepisodes of the episode 
$f8\rightarrow g8 \rightarrow d8$ versus age of culture in days. See text for 
explanation}
\label{fig:potter-conf}
\end{figure}

Fig.~\ref{fig:potter-norm-freq} shows a few such serial episodes disovered 
in the data from one of the cultures. (We have used inter-event time constraint of 
0--5 milli sec). The figure plots the frequency (in terms of the 
number of non-overlapping occurances as a fraction of the data length) for the frequent 
serial episodes versus the day on which the data is collected. In the figure, c5, e5, e6, 
d6 etc. are the pin numbers in the multi-electrode array which will be event types for the 
data mining algorithms. The increasing trend of the frequency is very clear and it is 
highly plausible that these episodes represent some underlying microcircuits that are 
developing as the culture ages. Fig.~\ref{fig:potter-conf} shows some further analysis 
of one of the episodes discovered, namely, 
$f8\rightarrow g8\rightarrow d8$. The figure plots ratio of the 
frequencies of $f8\rightarrow g8$ to that of $f8$ and similarly for $g8 \rightarrow d8$ 
and $g8$. This ratio, which we call the confidence for the subepisode, gives an 
indication of the chance that the second neuron spikes given the first one spiked. If 
the episode is really due to a circuit, we expect this confidence to be high but not too 
high. (If a spike from electrode g8 always follows a spike from electrode f8, then it 
may be that a single axon is making contact with both electrodes). As can be seen from the 
figure, the confidence values steadily grow with age of the culture and reach a reasonably 
high value. The increasing trend of the confidence values matches well with the 
increasing trend of the frequency of the episode also thus indicating that some 
underlying structure is responsible for the repeated occurrance of this episode. 
Similar behaviour is seen in case of other episodes and also for other cultures, 
thus indicating that the frequent episodes discovered are most probably due to 
coordinated firing by some group of neurons due to some underlying structures in the 
culture. 

In this data, there is no ground truth available regarding connections and hence it 
is not possible to directly validate the discovered episodes. However, we can indirectly 
get some evidence that the episodes capture some underlying structure in the neural system 
by looking at the sets of episodes obtained from same culture on different days and from 
different cultures. We considered six cultures, namely, culture 2-1 to culture 2-6. For 
each culture we considered the data from the last five days, namely days 31 to 35. (As we 
have seen from earlier figures, the circuits seem to stabilize only in the last week). 
However, in our data set, for culture 2-4 there was no recording on day 34 
Thus we have 29 data sets such as 2-1-31 (meaning culture 2-1, day 31) and so on. 
From each culture on each day, we have 30 minutes of data recording spontaneous activity. 
From each data 
set, we have taken a 10 minute duration data slice. 
From each such data 
slice, we identified top twenty most frequent 7-node serial episodes with inter-event 
interval constraint of 0--5 milli sec. (We want to consider long episodes because, as we 
saw earlier, it is highly unlikely to have large size frequent episodes by chance. The 
size of 7 is chosen so that all data sets have atleast twenty episodes of that size). Now 
we want to compare the sets of episodes discovered from different data slices. For this we 
need a measure of similarity between sets of episodes. 

We define a similarity score for two sets, $A, B$,  of episodes of size, say, N, as follows. 
We first count the 
number of N-node episodes that are common in the two sets and remove all the common 
ones from both sets. Then we replace each episode (in each set) with the two (N-1)-node 
subepisodes  obtained by dropping the first or last nodes in the original episode. We now 
count the common (N-1)-node episodes (in the two sets) and remove them. We go on like this, 
by replacing the left-over episodes with subepisodes of size one less and counting the 
common ones, till we reach episodes of size 1. Let $n_i$ denote the number of common 
episodes of size $i$. Then the similarity between the sets $A$ and $B$ is defined as 
\[\mbox{Sim}(A, B) = \sum_{i=1}^N \; 2^i n_i.\]
Since we want to view episodes as representing connections, similarity has to capture 
how much of the paths represented by different episodes are common. The above measure 
does just that and gives higher weightage to common long episodes. 

\begin{figure}[!htb]
\centering
\epsfig{file=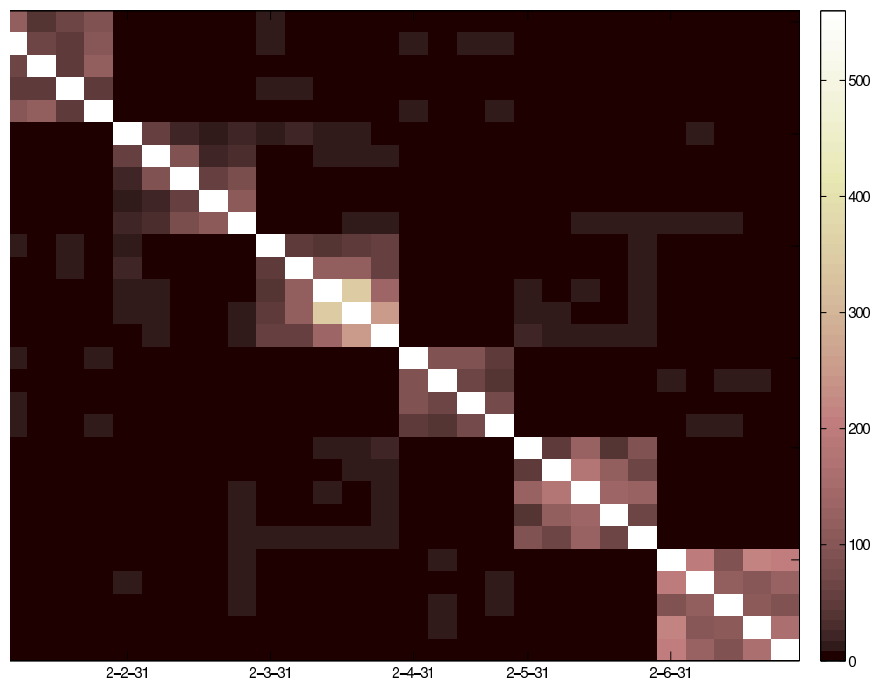,height=3.5in,width=4.5in}
\caption{Cross similarity between sets of frequent Serial Episodes. Each point 
represents similarity score between two sets of frequent episodes obtained from 
 10 minute duration data from pairs of culture-day combinations. Since the 
two axes are ordered in reverse orders, the reverse diagonal corresponds to 
the case where the two sets of frequent episodes are identical. 
The colour code is explained by the legend on the 
right. Note the high similarity scores around the reverse diagonal showing 
that sets of episodes from the same culture are much more similar than those from 
different cultures.}
\label{fig:potter-sim-matrix1}
\end{figure}

Fig.~\ref{fig:potter-sim-matrix1} shows the cross-similarity between the 
 sets of frequent episodes from the 29 data slices 
by colour coding similarity values. The axes indicate the culture-day 
combinations.  Note that the two axes are 
ordered differently so that the reverse diagonal represents similarity between 
identical sets of episodes. That is why the reverse diagonal has highest 
similarity values.  What is  
interesting is that data slices from the same culture but from different days are  
highly similar. This can be seen by observing the $5 \times 5$ submatrices 
around the reverse diagonal in the figure. (For the 2-4 culture, this submatrix 
is only $4 \times 4$ because there is no data for 2-4-34). 
This is in sharp contrast to the fact (as seen from the figure) that 
sets of episodes obtained from different cultures have very low similarity. These results 
strongly support the view that the frequent episodes capture some underlying structure in 
the neural system. 

\begin{figure}[!htb]
\centering
\epsfig{file=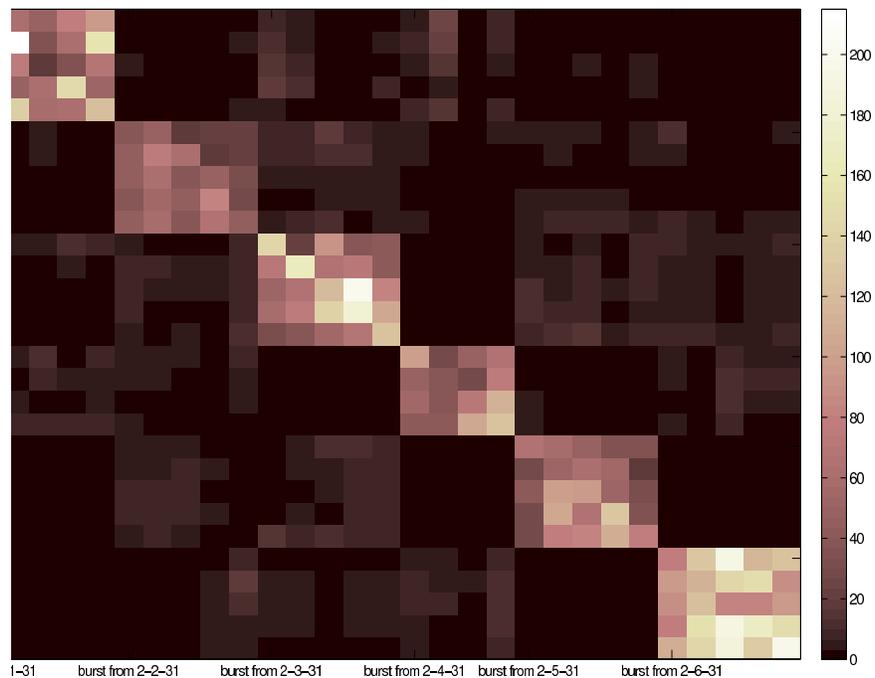,height=3.5in,width=4.5in}
\caption{Similarity between sets of frequent Serial Episodes. Each point 
represents similarity score between two sets of frequent episodes, one obtained from 
a 10 minute duration data and the other from a single burst, 
corresponding to pairs of culture-day combinations. Since the 
two axes are ordered in reverse orders, the reverse diagonal corresponds to 
same culture-day combinations. The colour code is explained by the legend on the 
right. Note the high similarity scores around the reverse diagonal showing 
 that sets of episodes from the same culture are much more similar than those from 
different cultures.}
\label{fig:potter-sim-matrix2}
\end{figure}

As said earlier, in the data we are considering, all cultures show very strong network-wide 
bursting activity that keeps occurring again and again. It is observed that most of our 
long episodes occur only during the burst period. Hence, an interesting question to ask 
is how far are individual bursts characterize the underlying system. For this, we do a 
similar analysis as above. As earlier, we obtain
 29 sets of frequent episodes by taking ten minute data from each culture on each of days 
31 to 35. We then get another 29 sets of frequent episode from 
data corresponding to a single burst (taken outside the ten minute duration) from each 
culture on each of the days 31 to 35. We then obtain similarities between these sets of 
frequent episodes and the results are shown in fig.~\ref{fig:potter-sim-matrix2}. As can be 
seen, the results are strikingly similar to the earlier case. Frequent episodes obtained 
from different cultures are highly dissimilar while those from the same culture but from 
different days are much more similar. The main difference here as compared to the earlier 
case, is along the reverse diagonal in the figure, where, while the similarity values 
are high, they are not the highest as in the earlier case. 
This is because (for the points along the reverse diagonal), one set of frequent 
episodes is obtained from data of ten minute duration whereas the other set is obtained 
from only a single burst which is typically only a couple of seconds long. In spite of 
this, these two sets of episodes show good similarity if they are from the same 
culture. This once again supports the view that there are some characteristic structures 
that are different for different cultures (which is natural because the synapses that form 
in a culture are mostly random) and these are captured well by our frequent episodes.

The results presented in figures \ref{fig:potter-sim-matrix1} and 
\ref{fig:potter-sim-matrix2} strongly indicate that the set of frequent episodes seem to 
characterize the activity of a neural culture well and hence they capture the 
underlying microcircuits which are different for different cultures.

\section{DISCUSSION}
\label{sec:dis}
In this paper we have considered the problem of analyzing multi-neuronal spike data 
sequences. We argued that the temporal data mining framework of frequent episode discovery 
is a very useful formalism for addressing this problem. We have shown how the 
structure of episodes with additional temporal constraints can capture most of the patterns 
that are of interest in this area of neurobiology. We have presented algorithms 
for discovering such frequent episodes and iluustrated the performance of the algorithms 
through simulations. We have considered both synthetic data generated through a 
realistic neuronal spike simulator as well as two sets of multi-neuronal data -- one 
obtained through Calcium imaging and the other obtained from multi-electrode array 
experiments. We have also presented extensive empirical results to show that our 
frequent episodes represent statistically significant patterns of correlated firings 
in the underlying neural system.

Analyzing multi-neuron spike data is a challanging problem of much current interest in 
neuroscience. Due to the abundance of experimental techniques one can now obtain data 
representing the simultaneous activity of many neurons grown in vivo. Thus algorithms that 
can unearth significant patterns in the data would go a long way in allowing neurobiologists 
to study firing patterns and microcircuits in neural assemblies. Such an understanding 
of the behaviour of interacting neurons is very useful in understanding vital issues 
such as learning and memory as well as for concrete applications such as brain computer 
interfaces. 

We have shown how one can detect many coordinated firing patterns such as order, 
synchrony as well as synfire chains in terms of episodes with appropriate temporal 
constraints. We illustrated the effectiveness of the algorithms by analyzing synthetically 
generated spike sequences that have embedded patterns in them. For this we have modelled 
each neuron as an inhomogeneous Poisson process whose spiking rate gets modified in 
response to the input received from other neurons. By building an interconnected  system 
od such neurons with some specific large excitatory connections along with many random 
connections we can embed different patterns in the system that is generating the spike 
data. Since the ground truth is known in these sequences, they serve as a useful test 
bed for assessing the capabilities of our algorithms. We have shown that our algorithms 
unerringly discover the underlying structure and also that they scale well even if there 
are multiple such patterns in the data. 

We have also used our neuronal spike data simulator to show that the kind of episodes we 
are taking about do not happen by chance. We have generated many sets of random data 
of both independent and dependent neurons and by using the `jitter method' 
\cite{date2001} idea of sampling from a null hypothesis are able to show that the maximum 
frequency of episodes in noise data are orders of magnitude smaller than minimum 
frequencies of relevant episodes when data contains patterns. This, we feel, is a very 
significant contribution of this paper because the null hypothesis we consider here 
goes far beyond the usual one of independent homogeneous Poisson processes. 

We have also illustrated the effectiveness of our method in analyzing data obtained 
from multi-neuronal systems grown in vivo. These results also show that, unlike the 
methods based on correlations, the data mining techniques proposed here are much more 
promising for getting information regarding the connectivity patterns. 

The data mining techniques we proposed here are much more efficient that the current methods 
based on analyzing correlations between spike trains and are also seen to be much more 
effective un unearthing interesting patterns that are relevant to understanding 
connectivity patterns. However, the full data analysis problem is very challenging because the 
spike trains are noisy stochastic processes where the useful patterns of coordinated 
activity can often be submerged by vast amount of background spiking. Thus, we view the 
results of this paper more as indicative of what the data mining approach can offer in 
this area rather than as solving the problem. Much more work is needed to develop these 
techniques to a level where they can become a routine tool for neurobiologists. We hope that 
this paper would contribute towards developing the necesary collaborations between 
neurobiologits and data mining researchers for such a fruitful activity.

\bibliographystyle{ieeetr}
\bibliography{spike-train}

\appendix
\section{Pseudo-code listing for Algorithms in the paper}
\begin{algorithm}[!htb]
\caption{\label{alg:count-parallel-EXPIRY}Non-overlapped count for parallel
episodes with expiry time constraint}
\begin{algorithmic}[1]
\REQUIRE Set $C$ of candidate $N$-node parallel episodes, event
streams $s=\langle(E_{1},t_{1}),\ldots,(E_{n},t_{n})\rangle$, frequency
threshold $\lambda_{min}\in[0,1]$, expiry time $T_{x}$
\ENSURE The set $F$ of frequent serial episodes in $C$

\FORALL {event types $A$}
	\STATE $waits(A)=\phi$
\ENDFOR

\FORALL {$\alpha\in C$}
	\STATE Initialize $autos(\alpha)=\phi$
\ENDFOR

\FORALL {$\alpha\in C$}
	\FORALL {event types $A\in\alpha$}
		\STATE Create node $s$ with $s.episode=\alpha$; $s.init=0$ ;
		\STATE $s.count=1$
		\STATE Add $s$ to $waits(A)$ 
		\STATE Add $s$ to $autos(\alpha)$
	\ENDFOR
	\STATE Set $\alpha.freq=0$
	\STATE Set $\alpha.counter=0$
\ENDFOR

\FOR {$i=1$ to $n$}
	\FORALL {$s\in waits(E_{i})$}
		\STATE Set $\alpha=s.episode$
		\STATE Set $j=s.count$ 
		\IF {$j>0$}
			\STATE Set $s.count=j-1$
			\STATE $\alpha.counter=\alpha.counter+1$
		\ENDIF
		\STATE $s.init=t_{i}$
		\STATE \{Expiry check\}
		\IF {$\alpha.counter=N$}
			\FORALL {$q\in autos(\alpha)$}
				\IF {$(t_{i}-q.init)>T_{x}$}
					\STATE $\alpha.counter=\alpha.counter-1$
					\STATE $q.count=q.count+1$
				\ENDIF
			\ENDFOR
		\ENDIF
		\STATE \{Update episode count\}
		\IF {$\alpha.counter=N$}
			\STATE Update $\alpha.freq=\alpha.freq+1$
			\STATE Reset $\alpha.counter=0$
			\FORALL {$q\in autos(\alpha)$}
				\STATE Update $q.count=1$
			\ENDFOR
		\ENDIF
	\ENDFOR
\ENDFOR
\STATE Output $F=\{\alpha\in C$ such that $\alpha.freq\ge n\lambda_{min}\}$
\end{algorithmic}
\end{algorithm}

\begin{algorithm}[!htb]
\caption{\label{alg:count-serial-INTERVAL-discovery}Non-overlapped serial
episodes count with inter-event interval constraints}
\begin{algorithmic}[1]

\REQUIRE Set $C$ of candidate $N$-node parallel episodes, event
streams $s=\langle(E_{1},t_{1}),\ldots,(E_{n},t_{n})\rangle$, frequency
threshold $\lambda_{min}\in[0,1]$, expiry time $T_{X}$
\ENSURE The set $F$ of frequent serial episodes in $C$
\FORALL{event types $A$}
	\STATE Initialize $waits(A)=\phi$
\ENDFOR
\FORALL{$\alpha\in C$}
	\STATE Set $prev=\phi$
	\FOR{$i=1$ to $N$} 
		\STATE Create $node$ with $node.visited=false$;
		$node.episode=\alpha$;
		$node.index=i$;
		$node.prev=prev$;
		$node.next=\phi$
		\IF{$i=1$}
			\STATE Add $node$ to $waits(\alpha[1])$ 
		\ENDIF
		\IF{$prev\ne\phi$}
			\STATE $prev.next=node$
		\ENDIF
	\ENDFOR
\ENDFOR
\FOR{$i=1$ to $n$}
	\FORALL{$node\in waits(E_{i})$}
		\STATE Set $accepted=false$
		\STATE Set $\alpha=node.episode$
		\STATE Set $j=node.index$ 
		\STATE Set $tlist=node.tlist$
		\IF{$j<N$}
			\FORALL{$tval\in tlist$}
				\IF{$(t_{i}-tval.init)>\alpha.t_{high}[j]$}
					\STATE Remove $tval$ from $tlist$
				\ENDIF
			\ENDFOR
		\ENDIF
		\IF{$j=1$}
			\STATE Update $accepted=true$
			\STATE Update $tval.init=t_{i}$
			\STATE Add $tval$ to $tlist$
			\IF{$node.visited=false$}
				\STATE Update $node.visited=true$
				\STATE Add $node.next$ to $waits(\alpha[j+1])$ 
			\ENDIF
		\ELSE
			\FORALL{$prev\_tval\in node.prev.tlist$}
				\IF{$t_{i}-prev\_tval\in(\alpha.t_{low}[j-1],\alpha.t_{high}[j-1]]$}
					\STATE Update $accepted=true$
					\STATE Update $tval.init=t_{i}$
					\STATE Add $tval$ to $tlist$
					\IF{$node.visited=false$}
						\STATE Update $node.visited=true$
						\IF{$node.index\le N-1$}
							\STATE Add $node.next$ to $waits(\alpha[j+1])$
						\ENDIF
					\ENDIF
				\ELSE
					\IF{$t_{i}-prev\_tval>\alpha.t_{high}[j-1]$}
						\STATE Remove $prev\_tval$ from $node.prev.tlist$
					\ENDIF
				\ENDIF
			\ENDFOR
		\ENDIF
		\IF{$accepted=true$ and $node.index=N$}
			\STATE Update $\alpha.freq=\alpha.freq+1$
			\STATE Set $temp=node$
			\WHILE{$temp\ne\phi$}
				\STATE Update $temp.visited=false$
				\IF{$temp.index\ne1$}
					\STATE Remove $temp$ from $waits(\alpha[temp.index])$
				\ENDIF
				\STATE Update $temp=temp.next$
			\ENDWHILE
		\ENDIF
	\ENDFOR
\ENDFOR
\STATE Output $F=\{\alpha\in C$ such that $\alpha.freq\ge n\lambda_{min}\}$
\end{algorithmic}
\end{algorithm}

\end{document}